\documentclass[thmsa,letterpaper]{article}
\usepackage{sw20lart}

\input{tcilatex}
\begin{document}

\author{S. Manoff\thanks{%
Work supported in part by the National Science Foundation in Bulgaria under
grant No. F-103} \\
Bulgarian Academy of Sciences\\
Institute for Nuclear Research and Nuclear Energy\\
Blvd. Tzarigradsko chaussee 72\\
1784 Sofia - Bulgaria}
\title{Kinematics of vector fields }
\date{Published in \textit{Complex structures and vector fields.}{\normalsize \ }%
World Scientific. Singapore - New Jersey - London - Hong Kong 1995, pp.
61-113}
\maketitle

\begin{abstract}
Different (not only by sign) affine connections are introduced for
contravariant and covariant tensor fields over a differentiable manifold by
means of a non-canonical contraction operator, defining the notion dual
space and commuting with the covariant and with the Lie-differential
operator. Classification of the linear transports on the basis of the
connections between the connections is given. Notion of relative velocity
and relative acceleration for vector fields are determined. By means of
these kinematic characteristics several other types of notions as shear
velocity, shear acceleration, rotation velocity, rotation acceleration,
expansion velocity and expansion acceleration are introduced and on their
basis the auto-parallel vector fields are classified.
\end{abstract}

\tableofcontents

\section{Introduction}

The evolution of the differential geometry is due to a great extent to ideas
connected with attempts for describing different types of physical
interactions by means of differential geometric methods. The created at the
beginning of the 20-th century theory of relativity carried out the
hypotheses of some geometers about connections between space-time and
material systems, evaluating in it, as well as ideas of many physicists,
trying to investigate mathematical models of physical systems by means of
differential-geometric structures (Lichnerowicz 1979).

The evolution of the special and the general theory of relativity and the
attempts for their generalization and connection with other theories of
physical interactions provided opportunity for using new geometrical
structures (different types of fiber bundles, geometries, different from the
Riemannian geometry, complex manifolds, different basic vector fields and
metric tensor fields, different connections) (Ivanenko, Pronin,
Sardanashvily 1985), (Barvinskii, Ponomariev, Obukhov 1985), (Hehl 1966,
1970, 1973, 1974). Different methods are also used in finding solutions of
equations for the gravitational field connected with differential-geometric
structures over manifolds (special vector and tensor fields, spinor fields
etc.). Problems, arising in solving the equations of modern gravitational
theories, induced an evolution of new approaches to existing mathematical
models and created preconditions for working out new differential-geometric
methods (Kramer, Stephani, MacCallum, Herlt 1980), (Kramer, Stephani 1983)..

For a century only the mathematical models of the space-time went from the
Euclidean, Minkowskian and (pseudo)Riemannian space-time to more
sophisticated spaces with linear (affine) connection and metric (Hecht, Hehl
1991), (Hehl, von der Heyde 1973), (Hehl, Kerlik 1978). The generalization
of the Newton's theory of gravitation in the Einstein's theory of
gravitation (ETG) was an important step toward the use of two essential
differential-geometric objects in the gravitational theory: the metric,
which allows the definition of a distance between two points of a manifolds,
considered as a model of space-time and the affine connection, which allows
the transport of a geometric object from one point to another point of a
manifold and a comparison of two objects at one and the same point. In the
Riemannian geometry these two geometric objects are connected each other -
the Levi-Civita (symmetric) connection can be given by means of the
Riemannian metric. This was at the beginning the mathematical basis for the
ETG and its generalization in the range of the Riemannian geometry. But
later on the generalizations went on two different directions: in the first
one two different metrics over the one and the same manifold were introduced
(bi-metric theory of gravitation (Rosen 1973, 1974), (Logunov,
Mestvirishvili 1989)) and in the second - two different connections for the
tensor fields over a manifold were introduced (bi-connection theory of
gravitation (Tchernikov 1987, 1988, 1990)). For Riemannian spaces these two
directions came one into another. In the last few years new attempts are
made to revive the ideas of Weyl, (Edington 1925) and (Schroedinger 1950)
for using manifolds with independent affine connection and metric (spaces
with affine connection and metric or $(L_n,g)$-spaces) as a model of
space-time in a theory of gravitation (Hecht, Hehl 1991). In such spaces the
connection for co-tangent vector fields (as dual to the tangent vector
fields) differs from the connection for the tangent vector fields only by
sign. The last fact is due to the definition of dual vector spaces over
points of a manifold, which is a trivial generalization of the definition of
algebraic dual vector spaces from the multi linear algebra (Greub 1978),
(Efimov, Rosendorn 1974), (Greub, Halperin, Vanstone 1972, 1973), (Bishop,
Goldberg 1968), . The hole modern differential geometry is build on the one
hand as a rigorous logical structure having as one of its main assumption
the canonical definition for algebraic dual vector spaces (with equal
dimensions) (Choquet-Bruhat, DeWitte-Morette, Dillard-Bleik 1977). On the
other hand, the possibility of introducing a non-canonical definition for
algebraic dual vector spaces (with equal dimensions) has been pointed out by
many mathematicians (Kobayashi, Nomizu 1963) who have not exploited this
possibility for further evolution of the differential-geometric structures
and its applications. The canonical definition of dual spaces is so
naturally embedded in the ground of the differential geometry that no need
has occurred for changing it (Matsushima 1972), (Boothby 1975), (Lovelock,
Rund 1975), (Norden 1976). But the last time evolution of the mathematical
models for describing the gravitational interaction on classical level shows
a tendency to generalizations using spaces with affine connection and
metric, which can be also generalized using the freedom of the
differential-geometric preconditions. The fact, that affine connection,
which in a point or over a curve in Riemannian spaces can vanish (principle
of equivalence in ETG), can also vanish under special choice of the basic
system in a space with affine connection and metric (von der Heyde 1975),
(Iliev 1992), shows that the equivalence principle in the ETG is only a
corollary of the mathematical apparatus used in this theory. Therefore,
every differentiable manifold with affine connection and metric can be used
as a model for space-time in which the equivalence principle holds. But, if
the manifold has two different (not only by sign) connections for tangent
and co-tangent vector fields, the situation changes and is worth being
investigated.

In Section 1. the notions contravariant and covariant affine connection are
defined for contravariant and covariant tensor fields over differentiable
manifolds. It is shown that these two different (not only by sign)
connections can be introduced by means of changing the canonical definition
of dual vector spaces (respectively of dual vector fields).

In Einstein's theory of gravitation (ETG) kinematic notions related to the
notion relative velocity such as shear velocity tensor (shear velocity,
shear) $\sigma $, rotation velocity tensor (rotation velocity, rotation) $%
\omega $ and expansion velocity (expansion) $\theta $, are used in finding
solutions of special types of Einstein's field equations and in the
description of the properties of the (pseudo)Riemannian spaces without
torsion ($V_n$-spaces). By means of these notions a classification of $V_n$%
-spaces, admitting special types of geodesic vector fields has been proposed
(Ehlers 1961). The same kinematic characteristics are also necessary for
description of the projections of the Riemannian (curvature) tensor and the
Ricci tensor along a non-isotropic (non-null) vector field (Kramer,
Stephani, MacCallum, Herlt 1980) and in obtaining and using the Raychaudhuri
identity (Hawking, Ellis 1973) in $V_n$-spaces.

The kinematic characteristics, connected with the notion relative velocity
can be generalized for vector fields over differentiable manifolds with
contravariant and covariant affine connection and metric (($\overline{L}_n,g$%
)-spaces) so that in the case of ($L_n,g$)- and $V_n$-spaces (as a special
case of ($\overline{L}_n,g$)-spaces) and for normalized non-isotropic vector
fields these characteristics are the same as those, introduced in the ETG.
In analogous way as in the case of the kinematic characteristics, related to
the notion of relative velocity, it is possible to introduce kinematic
characteristics, related to the notion of relative acceleration such as
shear acceleration tensor (shear acceleration), rotation acceleration tensor
(rotation acceleration) and expansion acceleration (Manoff 1985, 1992).

In Section 2., 3. and 4. the corresponding for ($\overline{L}_n,g$)-spaces
notions of relative velocity and relative acceleration are introduced. By
means of these kinematic characteristics several other types of notions such
as shear velocity, shear acceleration, rotation velocity, rotation
acceleration, expansion velocity and expansion acceleration are investigated
and on their basis the auto-parallel vector fields in ($\overline{L}_n,g$%
)-spaces are classified. The generalizations compared with those in ($L_n,g$%
)-spaces (differentiable manifolds with affine connection and metric) appear
only in the explicit forms of the expressions, written in a corresponding
basis (or in other words - only in index forms).

On the basis of the introduced notions deviation equations (playing
important role in gravitational physics) and Lagrangian theories of tensor
fields can be considered in ($\overline{L}_n,g)$-spaces. If kinematic
characteristics of a dynamic system are given as preconditions, then the
corresponding type of differentiable manifold (which allows such
characteristics) can be chosen as a model of space-time, where the evolution
of the system is taking place. This idea connects different
differential-geometric structures used for describing physical systems on
classical level. The main objects taken in such type of considerations can
be given schematically as follows

$
\begin{array}{c}
\\ 
\frame{$
\begin{array}{c}
\text{Differentiable manifolds} \\ 
\text{with contravariant and covariant affine connections and metric}
\end{array}
$} \\ 
\\ 
\fbox{$
\begin{array}{c}
\text{Kinematic characteristics} \\ 
\text{of contravariant vector fields}
\end{array}
$} \\ 
\\ 
\fbox{$
\begin{array}{c}
\text{Deviation equations} \\ 
\text{for vector fields}
\end{array}
$} \\ 
\\ 
\fbox{Lagrangian theory of tensor fields} \\ 
=.=
\end{array}
$

In the present paper we will concentrate our attention only on the kinematic
characteristics of contravariant vector fields but some main ideas and
definitions will be outlined.

\section{Differentiable manifolds with contravariant and covariant affine
connection and metric [$(\overline{L}_n,g)$-spaces]}

The notion algebraic dual vector space can be introduced in such a way
(Efimov, Rosendorn 1974), in which the two vector spaces (the considered and
its dual vector space) are two independent (finite) vector spaces with equal
dimensions.

\subsection{Contraction operator}

Let $X$ and $X^{*}$ be two vector spaces with equal dimensions $\dim X=\dim
X^{*}=n.$ Let $S$ be an operator (mapping) such that to every pair of
elements $u\in X$ and $p\in X^{*}$ sets an element of the field $K$ $(R$ or $%
C)$, i.e. 
\[
S:(u,p)\rightarrow z\in K\text{ , }u\in X\text{ , }p\in X^{*}\text{ .} 
\]

\textbf{Definition 1.} The operator (mapping) $S$ is called \textit{%
contraction operator }$S$, if it is a bi-linear symmetric mapping, i.e. if
it fulfills the following conditions:

a) $S(u,p_1+p_2)=S(u,p_1)+S(u,p_2)$ , $\forall u\in X$ $,\forall p_i\in
X^{*} $ , $i=1,2$ ,

b) $S(u_1+u_2,p)=S(u_1,p)+S(u_2,p)$ , $\forall u_i\in X$ , $i=1,2$ , $%
\forall $$p\in X^{*}$ ,

c) $S(\alpha u,p)=S(u,\alpha p)=\alpha .S(u,p)$ , $\alpha \in K$ ,

d) non-degeneracy: if $u_1,...,u_n$ are linear independent in $X$ and $%
S(u_1,p)=0$, ... , $S(u_n,p)=0$, then the $p$ is the null element in $X^{*}$
. In analogous way, if $p_1$,..., $p_n$ are linear independent in $X^{*}$
and $S(u,p_1)=0$, ... , $S(u,p_n)=0$, then $u$ is the null element in $X$,

e) symmetry: $S(u,p)=S(p,u)$ , $\forall u\in X$ , $\forall p\in X^{*}$ .

Let $e_1,...,e_n$ be an arbitrary basis in $X,$ and let $e^1,...,e^n$ be an
arbitrary basis in $X^{*}$ . Let $u=u^i.e_i\in X$ and $p=p_k.e^k\in X^{*}.$

From the properties a) and b) it follows that 
\[
S(u,p)=f^k\text{ }_i.u^i.p_k=u^i.p_{\overline{i}}=p_k.u^{\overline{k}}\text{
, }p_{\overline{i}}=f^k\text{ }_i.p_k\text{ , }u^{\overline{k}}=f^k\text{ }%
_i.u^i\text{ ,} 
\]

\noindent where 
\[
f^k\text{ }_i=S(e_i,e^k)=S(e^k,e_i)\in K\text{ .} 
\]

In this way, the result of the action of the contraction operator $S$ is
expressed in terms of a bi-linear form. The property non-degeneracy d) means
the non-degeneracy of the bi-linear form. The result $S(u,p)$ can be defined
in different ways by giving arbitrary numbers $f^k$ $_i\in K,$ for which the
condition $\det (f^k$ $_i)\neq 0$ and the conditions a) - d) are fulfilled.

\textbf{Remark.} In the canonical approach $S=C$ and $%
C(e_i,e^k)=C(e^k,e_i)=g_i^k$ , $g_i^k=1$ for $k=i,$ $g_i^k=0$ for $i\neq k$.
The contraction operator $C$ is the corresponding to the canonical approach
mapping(Boothby 1975), (Matsushima 1972) 
\[
C(u,p)=C(p,u)=p(u)=p_i.u^i\text{ .} 
\]

\textbf{Definition 2.} (Mutually) \textit{algebraic dual vector spaces} :=
The spaces $X$ and $X^{*}$ are called (mutually) dual spaces, if an
contraction operator acting on them is given and they are considered
together with this operator (i.e. $(X,X^{*},S)$ with $\dim X=n=\dim X^{*}$
defines the two (mutually) dual vector spaces $X$ and $X^{*}$).

\textbf{Remark.} The generalization of the notion of algebraic dual vector
spaces for the case of vector fields over differentiable manifold is a
trivial one. The vector fields are considered as sections of vector bundles
over a manifold. The vector bases become dependent on the points of the
manifold and the numbers $f^i$ $_j$ are considered as functions over the
manifold.

Vector and tensor fields over a differentiable manifold are provided with
the structure of a linear (vector) space by defining the corresponding
operations at every point of the manifold.

Thus, the definition of algebraic dual vector spaces over manifolds by means
of the contraction operator $S$ as a generalization of the contraction
operator $C$ allows considerations including functions $f^i$ $_j\in C^k(M)$
instead of the Kroneker symbol $g_j^i$.

\subsection{Covariant differential operator. Contravariant and covariant
affine connection}

The notion affine connection can be defined in different ways but in all
definitions a linear mapping is given, which to a given vector of a vector
space over a point $x$ of a manifold $M$ juxtaposes a corresponding vector
from the same vector space in this point. The corresponding vector is
identified as vector of the vector space over another point of the manifold $%
M$. The way of identification is called \textit{transport} from a point to
another point of the manifold.

\textbf{Definition 3.}\textit{\ Affine connection} over a differentiable
manifold $M$. Let $V(M)$ $(\dim M=n)$ be the set of all (smooth) vector
fields over the manifold $M$. The mapping 
\[
\nabla :V(M)\times V(M)\rightarrow V(M)\text{ ,} 
\]

\noindent by means of 
\[
\nabla (u,v)\rightarrow \nabla _uv\text{ , }u,v\in V(M)\text{ ,} 
\]

\noindent with the following properties

a) $\nabla _u(v+w)=\nabla _uv+\nabla _uw$ , $u,v,w\in V(M)$ ,

b) $\nabla _u(fv)=(uf).v+f.\nabla _uv$ , $f\in C^r(M)$ , $r\geq 1$ ,

c) $\nabla _{u+v}w=\nabla _uw+\nabla _vw$ ,

d) $\nabla _{fu}v=f.\nabla _uv$ ,

\noindent is called \textit{affine connection} over the manifold $M$.

\textbf{Definition 4.}\textit{\ Covariant differential operator.} The linear
differential operator (mapping) $\nabla _u$ with the following properties

a) $\nabla _u(v+w)=\nabla _uv+\nabla _uw$ , $u,v,w\in V(M)$ ,

b) $\nabla _u(fv)=(uf).v+f.\nabla _uv$ , $f\in C^r(M)$ , $r\geq 1$ ,

c) $\nabla _{u+v}w=\nabla _uw+\nabla _vw$ ,

d) $\nabla _{fu}v=f.\nabla _uv$ ,

e) $\nabla _uf=uf$ , $f\in C^r(M)$ , $r\geq 1$ ,

f)$\nabla _u(v\otimes w)=\nabla _uv\otimes w+v\otimes \nabla _uw$ (Leibnitz
rule), $\otimes $ is the sign for tensor product,

\noindent is called \textit{covariant differential operator along the vector 
}$u $.

The result $\nabla _uv$ of the action of the covariant differential operator
on $v$ is often called \textit{covariant derivative of the vector field }$v$
along the vector field $u$.

In a given chart (co-ordinate system) the determination of $\nabla
_{e_\alpha }e_\beta $ in the basis $\{e_\alpha \}$ defines the components $%
\nabla _{\beta \gamma }^\alpha $ of the affine connection $\nabla $%
\[
\nabla _{e_\alpha }e_\beta =\nabla _{\alpha \beta }^\gamma .e_\gamma \text{
, }\alpha ,\beta ,\gamma =1,...,n\text{ .} 
\]

\textbf{Definition 5.}\textit{\ Space with affine connection.}
Differentiable manifold $M,$ provided with affine connection $\nabla $, i.e.
the pair $(M,\nabla ),$ is called space with affine connection.

The action of the covariant differential operator on a contravariant
(tangential) co-ordinate basic vector field $\partial _i$ over $M$ along
another contravariant co-ordinate basic vector field is determined by the
affine connection $\nabla =\Gamma $ with the components $\Gamma _{ij}^k$ in
a given chart (co-ordinate system) defined through 
\[
\nabla _{\partial _j}\partial _i=\Gamma _{ij}^k.\partial _k\text{ .} 
\]

For a non-co-ordinate contravariant basis 
\[
e_\alpha \text{ (or }e_i\text{)}\in T(M),T(M)=\cup _{x\in M}T_x(M) 
\]
\[
\nabla _{e_\beta }e_\alpha =\Gamma _{\alpha \beta }^\gamma .e_\gamma \text{ .%
} 
\]

\textbf{Definition 6. }\textit{Contravariant affine connection.} Affine
connection $\nabla =$$\Gamma $ induced by the action of the covariant
differential operator on contravariant vector fields is called contravariant
affine connection.

The action of the covariant differential operator on a covariant (dual to
contravariant) basic vector field $e^\alpha $ ($e^\alpha \in T^{*}(M)$ , $%
T^{*}(M)=\cup _{x\in M}T_x^{*}(M))$ along a contravariant basic
(non-co-ordinate) vector field $e_\beta $ is determined by affine connection 
$\nabla =P$ with components $P_{\beta \gamma }^\alpha $ defined through 
\[
\nabla _{e_\beta }e^\alpha =P_{\gamma \beta }^\alpha .e^\gamma \text{ .} 
\]

For a co-ordinate basis $dx^i$%
\[
\nabla _{\partial _j}dx^i=P_{kj}^i.dx^k\text{ .} 
\]

\textbf{Definition 7.} \textit{Covariant affine connection.} Affine
connection $\nabla =P$ induced by action of the covariant differential
operator on covariant vector fields is called covariant affine connection.

\textbf{Definition 8.}\textit{\ Space with contravariant and covariant
affine connection (}$\overline{L}_n$-space). Differentiable manifold
provided with contravariant affine connection $\Gamma $ and covariant affine
connection $P$ is called space with contravariant and covariant affine
connection.

\textbf{Definition 9.}\textit{\ Space with contravariant and covariant
affine connection, and metric (}$(\overline{L}_n,g)$-space). Differentiable
manifold provided with contravariant affine connection $\Gamma $ and
covariant affine connection $P,$ and metric $g$ is called space with
contravariant and covariant affine connection and metric.

The connection between the two connections $\Gamma $ and $P$ is based on the
connection between the two dual spaces $T(M)$ and $T^{*}(M)$, which on its
side is based on the existence of the contraction operator $S$. Usually
commutation relations are required between the contraction operator and the
covariant differential operator in the form 
\[
S\circ \nabla _u=\nabla _u\circ S\text{ .} 
\]

If the last operator equality in the form $\nabla _{e_\gamma }\circ S=S\circ
\nabla _{e_\gamma }$ is used for acting on the tensor product $e^\alpha
\otimes e_\beta $ of two basic vector fields $e^\alpha \in T^{*}(M)$ and $%
e_\beta \in T(M),$ then 
\[
\nabla _{e_\gamma }(S(e^\alpha \otimes e_\beta ))=S(\nabla _{e_\gamma
}(e^\alpha \otimes e_\beta ))\text{ ,} 
\]

\noindent and the relation follows 
\[
\begin{array}{c}
e_\gamma f^\alpha \text{ }_\beta =\Gamma _{\beta \gamma }^\delta .f^\alpha 
\text{ }_\delta +P_{\delta \gamma }^\alpha .f^\delta \text{ }_\beta \text{ , 
} \\ 
\text{(in a non-co-ordinate basis)}
\end{array}
\]

\noindent or 
\[
\begin{array}{c}
f^i\text{ }_{j,k}=\Gamma _{jk}^l.f^i\text{ }_l+P_{lk}^i.f^l\text{ }_j\text{ ,%
} \\ 
f^i\text{ }_{j,k}=\partial _kf^i\text{ }_j\text{ ,} \\ 
\text{(in a co-ordinate basis).}
\end{array}
\]

The last equality can be considered from two different points of view:

1. If $P_{jk}^i(x^l)$ and $\Gamma _{jk}^i(x^l)$ are given functions of
co-ordinates in $M$, then the equality appears as a system of equations for
the unknown functions $f_{.j}^i(x^k)$. The solutions of these equations
determine the action of the contraction operator $S$ on the basic vector
fields for given components of both the connections. The integrability
conditions for the equations can be written in the form 
\[
R^m\text{ }_{jkl}.f^i\text{ }_m+P^i\text{ }_{mkl}.f^m\text{ }_j=0\text{ ,} 
\]

\noindent where $R^m$ $_{jkl}$ are the components of the contravariant
curvature tensor, constructed by means of the contravariant affine
connection $\Gamma $, and $P_{.mkl}^i$ are the components of the covariant
curvature tensor, constructed by means of the covariant affine connection $P$%
, where $[R(\partial _i,\partial _j)]dx^k=P_{.lij}^k.dx^l$, $[R(\partial
_i,\partial _j)]\partial _k=R^l$ $_{kij}.\partial _l$ , $R(\partial
_i,\partial _j)=\nabla _{\partial _i}\nabla _{\partial _j}-\nabla _{\partial
_j}\nabla _{\partial _i}$.

2. If $f^i$ $_j(x^l)$ are given as functions of the co-ordinates in $M$,
then the conditions for $f^i$ $_j$ determine the connection between the
components of the contravariant affine connection $\Gamma $ and the
components of the covariant affine connection $P$ on the ground of the
predetermined action of the contraction operator $S$ on basic vector fields.

If $S=C$, i.e. $f^i$ $_j=g_j^i$ , then the conditions for $f^i$ $_j$ are
fulfilled for every $P=-\Gamma $, i.e. 
\[
P_{jk}^i=-\Gamma _{jk}^i\text{ .} 
\]

This fact can be formulated as the following proposition:

\textbf{Proposition 1. }$S=C$ is a sufficient condition for $P=-\Gamma $ $%
(P_{jk}^i=-\Gamma _{jk}^i)$.

\textbf{Corollary. }If $P\neq -\Gamma $, then $S\neq C$ , i.e. if the
covariant affine connection $P$ has to be different from the contravariant
affine connection $\Gamma $ not only by sign, then the contraction operator $%
S$ has to be different from the canonical contraction operator $C$ (if $S$
commutes with the covariant differential operator).

The corollary allows introduction of different (not only by sign)
contravariant and covariant connection by using contraction operator $S$,
different from the canonical operator $C$.

\textbf{Example.} If $f^i$ $_j=\varphi .g_j^i$, where $\varphi \in C^r(M)$, $%
\varphi \neq 0$, then $P_{jk}^i=-\Gamma _{jk}^i+(\log \varphi )_{,k}.g_j^i$ .

The covariant derivatives of contravariant vector fields can be written in
an arbitrary co-ordinate or non-co-ordinate basis 
\[
\begin{array}{c}
\nabla _uv=(v^i \text{ }_{,j}+\Gamma _{kj}^iv^k)u^j.\partial _i=v^i\text{ }%
_{;j}u^j.\partial _i\text{ , }u,v\in T(M)\text{ ,} \\ 
\text{(in a co-ordinate basis),} \\ 
\nabla _uv=(e_\beta v^\alpha +\Gamma _{\gamma \beta }^\alpha .v^\gamma
)u^\beta .e_\alpha =v^\alpha \text{ }_{/\beta }u^\beta .e_\alpha \text{ ,}
\\ 
\nabla _uv=(e_jv^i+\Gamma _{kj}^iv^k)u^j.e_i=v^i \text{ }_{/j}u^j.e_i\text{ ,%
} \\ 
\text{(in a non-co-ordinate basis with different type of indices).}
\end{array}
\]

In analogous way the covariant derivative of covariant vector fields can be
written in an arbitrary co-ordinate or non-co-ordinate basis 
\[
\begin{array}{c}
\nabla _up=(p_{i,j}+P_{ij}^kp_k)u^j.dx^i=p_{i;j}u^j.dx^i \text{ , }p\in
T^{*}(M)\text{ , }u\in T(M)\text{ ,} \\ 
\text{(in a co-ordinate basis),} \\ 
\nabla _up=(e_\beta p_\alpha +P_{\alpha \beta }^\gamma .p_\gamma )u^\beta
.e^\alpha =p_{\alpha /\beta }u^\beta .e^\alpha \text{ ,} \\ 
\nabla _up=(e_jp_i+P_{ij}^k.p_k)u^j.e^i=p_{i/j}u^j.e^i \text{ ,} \\ 
\text{(in a non-co-ordinate basis with different type of indices).}
\end{array}
\]

The action of the covariant differential operator on contravariant and
covariant tensor fields as well as on mixed tensor fields with rank $\succ 1$
is generalized in trivial manner on the ground of the Leibnitz rule, which
holds for this operator.

If the Kroneker tensor is defined in the form 
\[
Kr=g_j^i.\partial _i\otimes dx^j=g_\beta ^\alpha .e_\alpha \otimes e^\beta 
\text{ ,} 
\]

\noindent then the components of the contravariant and covariant affine
connection differ from each other by the components of the covariant
derivative of the Kroneker tensor, i.e. 
\[
\begin{array}{c}
\Gamma _{jk}^i+P_{jk}^i=g_{j;k}^i\text{ ,} \\ 
\Gamma _{\beta \gamma }^\alpha +P_{\beta \gamma }^\alpha =g_{\beta /\gamma
}^\alpha \text{ .}
\end{array}
\]

\textbf{Remark.} In the special case, when $S=C$, and in the canonical
approach $g_{j;k}^i=0$ ($g_{\beta /\gamma }^\alpha =0$).

\subsection{Lie-differential operator}

The Lie-differential operator $\pounds _\xi $ along the contravariant vector
field $\xi $ appears as another operator, which can be constructed by means
of contravariant vector field. His definition can be considered as a
generalization of the notion Lie derivative of tensor fields (Slebodzinski
1931), (Yano 1957), (Kobayashi, Nomizu 1963), (Lightman, Press, Price,
Teukolsky 1975).

\textbf{Definition 10. }$\pounds _\xi $ := \textit{Lie-differential operator
along the contravariant vector field }$\xi $ with the following properties:

a) $\pounds _\xi :V\rightarrow \overline{V}=\pounds _\xi V$ , $V,\overline{V}%
\in \otimes ^l(M)$ ,

b) $\pounds _\xi :W\rightarrow \overline{W}=\pounds _\xi W$ , $W,\overline{W}%
\in \otimes _k(M)$ ,

c) $\pounds _\xi :K\rightarrow \overline{K}=\pounds _\xi K$ , $K,\overline{K}%
\in \otimes ^l$ $_k(M)$ ,

d) linear operator with respect to tensor fields,

$\pounds _\xi (\alpha .V_1+\beta .V_2)=\alpha .\pounds _\xi V_1+\beta
.\pounds _\xi V_2$ , $\alpha ,\beta \in F(R$ or $C)$ , $V_i\in \otimes ^l(M)$
, $i=1,2$,

$\pounds _\xi (\alpha .W_1+\beta .W_2)=\alpha .\pounds _\xi W_1+\beta
.\pounds _\xi W_2$ , $W_i\in \otimes _k(M)$ , $i=1,2$,

$\pounds _\xi (\alpha .K_1+\beta .K_2)=\alpha .\pounds _\xi K_1+\beta
.\pounds _\xi K_2$ , $K_i\in \otimes ^l$ $_k(M)$ , $i=1,2$,

e) linear operator with respect to the contravariant field $\xi $ ,

$\pounds _{\alpha \xi +\beta u}=\alpha .\pounds _\xi +\beta .\pounds _u$ , $%
\alpha ,\beta \in F(R$ or $C),$ $\xi ,u\in T(M)$ ,

f) differential operator, obeying the Leibnitz rule,

$\pounds _\xi (S\otimes U)=\pounds _\xi S\otimes U+S\otimes \pounds _\xi U$
, $S\in \otimes ^m$ $_q(M)$ , $U\in \otimes ^k$ $_l(M)$ ,

g) action on function $f\in C^r(M)$ , $r\geq 1$ ,

$\pounds _\xi f=\xi f$ , $\xi \in T(M)$ ,

h) action on contravariant vector field,

$\pounds _\xi u=[\xi ,u]$ , $\xi ,u\in T(M)$ , $[\xi ,u]=\xi \circ u-u\circ
\xi $ ,

$\pounds _\xi e_\alpha =[\xi ,e_\alpha ]=-(e_\beta \xi ^\alpha -\xi ^\gamma
.C_{\gamma \beta }$ $^\alpha )e_\alpha $ ,

$\pounds _{e_\alpha }e_\beta =[e_\alpha ,e_\beta ]=C_{\alpha \beta }$ $%
^\gamma .e_\gamma $ , $C_{a\beta }$ $^\gamma \in C^r(M)$ ,

$\pounds _\xi \partial _i=-\xi ^j$ $_{,i}.\partial _j$ , $\pounds _{\partial
_i}\partial _j=[\partial _i,\partial _j]=0$ ,

i) action on covariant basic vector field,

$\pounds _\xi e^\alpha =k^\alpha $ $_\beta (\xi ).e^\beta $ , $\pounds
_{e_\gamma }e^\alpha =k^\alpha $ $_{\beta \gamma }.e^\beta $ ,

$\pounds _\xi dx^i=k^i$ $_j(\xi ).dx^j$ , $\pounds _{\partial _k}dx^i=k^i$ $%
_{jk}.dx^j$ .

The action of the Lie-differential operator on covariant basic vector field
is determined by its action on contravariant basic vector field and the
commutation relations between the Lie-differential operator and the
contraction operator $S$.

\subsubsection{Lie derivative of covariant co-ordinate basic vector fields}

The commutation relations between the Lie-differential operator $\pounds
_\xi $ and the contraction operator $S$ in the case of basic co-ordinate
vector fields can be written in the form 
\begin{equation}
\begin{array}{c}
\pounds _\xi \circ S(dx^i\otimes \partial _j)=S\circ \pounds _\xi
(dx^i\otimes \partial _j)\text{ ,} \\ 
\pounds _\xi \circ S(e^\alpha \otimes e_\beta )=S\circ \pounds _\xi
(e^\alpha \otimes e_\beta )\text{ ,}
\end{array}
\label{03.1}
\end{equation}

\noindent where 
\begin{equation}
\begin{array}{c}
\pounds _\xi \circ S(dx^i\otimes \partial _j)=\xi f^i\text{\thinspace }_j=f^i%
\text{ }_{j,k}\xi ^k\text{ ,} \\ 
S\circ \pounds _\xi (dx^i\otimes \partial _j)=S\circ (\pounds _\xi
dx^i\otimes \partial _j+dx^i\otimes \pounds _\xi \partial _j)= \\ 
=S(\pounds _\xi dx^i\otimes \partial _j)+S(dx^i\otimes \pounds _\xi \partial
_j)\text{ .}
\end{array}
\label{03.2}
\end{equation}

Since $\pounds _\xi dx^i=k^i$ $_j(\xi ).dx^j$ , $\pounds _{\partial
_k}dx^i=k^i$ $_{jk}.dx^j$, where $k^i$ $_j(\xi )$ $\in C^r(M)$, $k^i$ $%
_{jk}\in C^r(M)$, $k^i$ $_j(\xi )$ and $k^i$ $_{jk}$ have to be determined
by means of the commutation relations between $\pounds _\xi $ and $S$ and
their action on $dx^i$ and $\partial _j$ (respectively $e^\alpha $ and $%
e_\beta $) on the basis of the relations 
\begin{equation}
\begin{array}{c}
\pounds _\xi \partial _j=-\xi ^k\text{ }_{,j}\partial _k\text{ , }%
S(dx^i\otimes \pounds _\xi \partial _j)=-\xi ^k\text{ }_{,j}f^i\text{ }_k%
\text{ ,} \\ 
S(\pounds _\xi dx^i\otimes \partial _j)=k^i\text{ }_l(\xi ).f^l\text{ }_j%
\text{ ,} \\ 
S[\pounds _\xi (dx^i\otimes \partial _j)]=k^i\text{ }_l(\xi ).f^l\text{ }%
_j-\xi ^k\text{ }_{,j}.f^i\text{ }_k= \\ 
=\pounds _\xi [S(dx^i\otimes \partial _j)]=f^i\text{ }_{j,k}\xi ^k\text{ .}
\end{array}
\label{03.3}
\end{equation}

From the last expression the condition follows for $k^i$ $_l(\xi )$%
\begin{equation}
k^i\text{ }_l(\xi ).f^l\text{ }_j=\xi ^k\text{ }_{,j}.f^i\text{ }_k+f^i\text{
}_{j,k}\xi ^k\text{ .}  \label{03.4}
\end{equation}

By means of the non-degenerate inverse matrix $(f^i$ $_j)^{-1}=(f_j$ $^i)$
and the connections $f^i$ $_k.f_j$ $^k=g_j^i$ , $f^k$ $_i.f_k$ $^j=g_i^j$ ,
after multiplication of the equality for $k^i$ $_l(\xi )$ with $f_m$ $^j$
and summation over $j$, the explicit form for $k^i$ $_j(\xi )$ is obtained
in the form 
\begin{equation}
k^i\text{ }_j(\xi )=f_j\text{ }^l.\xi ^k\text{ }_{,l}.f^i\text{ }_k+f_j\text{
}^l.f^i\text{ }_{l,k}\xi ^k\text{ .}  \label{03.5}
\end{equation}

For $\pounds _{\partial _k}dx^i=k^i$ $_j(\partial _k).dx^j=k^i$ $_{jk}.dx^j$
it follows the corresponding form 
\begin{equation}
\begin{array}{c}
\pounds _{\partial _k}dx^i=k^i\text{ }_{jk}.dx^j=f_j\text{ }^l.f^i\text{ }%
_{l,k}.dx^j\text{ ,} \\ 
k^i\text{ }_{jk}=f_j\text{ }^l.f^i\text{ }_{l,k}\text{ .}
\end{array}
\label{03.6}
\end{equation}

On the other hand, from the commutation relations between $S$ and the
covariant differential operator $\nabla _\xi $ the connection between the
partial derivatives of $f^i$ $_j$ and the components of the contravariant
and covariant connections $\Gamma $ and $P$ follows in the form 
\begin{equation}
f^i\text{ }_{l,k}=P_{mk}^i.f^m\text{ }_l+\Gamma _{lk}^m.f^i\text{ }_m\text{ .%
}  \label{03.7}
\end{equation}

After substituting the last expression in the expressions for $k^i$ $_j(\xi
) $ and for $k^i$ $_{jk}$ the corresponding quantities are obtained in the
forms 
\begin{equation}
k^i\text{ }_j(\xi )=f_j\text{ }^l.\xi ^k\text{ }_{,l}.f^i\text{ }%
_k+(P_{jk}^i+f_j\text{ }^l.\Gamma _{lk}^m.f^i\text{ }_m)\xi ^k\text{ ,}
\label{03.8}
\end{equation}
\[
k^i\text{ }_j(\partial _k)=k^i\text{ }_{jk}=P_{jk}^i+f_j\text{ }^l.\Gamma
_{lk}^m.f^i\text{ }_m\text{ ,} 
\]

\begin{equation}
\pounds _\xi dx^i=[f_j\text{ }^l.\xi ^k\text{ }_{,l}.f^i\text{ }%
_k+(P_{jk}^i+f_j\text{ }^l.\Gamma _{lk}^m.f^i\text{ }_m).\xi ^k].dx^j\text{ ,%
}  \label{03.9}
\end{equation}
\begin{equation}
\begin{array}{c}
\pounds _{\partial _k}dx^i=k^i\text{ }_{jk}.dx^j= \\ 
=(P_{jk}^i+f_j\text{ }^l.\Gamma _{lk}^m.f^i\text{ }_m)dx^j\text{ .}
\end{array}
\label{03.10}
\end{equation}

If we introduce the abbreviations 
\begin{equation}
\xi _{\text{ }}^{\overline{i}}\text{ }_{,\underline{j}}=f^i\text{ }_k.\xi ^k%
\text{ }_{,l}f_j\text{ }^l\text{ , }\Gamma _{\underline{j}k}^{\overline{i}%
}=f_j\text{ }^l.\Gamma _{lk}^m.f^i\text{ }_m\text{ ,}  \label{03.11}
\end{equation}

\noindent then the Lie derivatives of covariant co-ordinate basic vector
fields $dx^i$ along the contravariant vector fields $\xi $ and $\partial _k$
can be written in the form 
\begin{equation}
\begin{array}{c}
\pounds _\xi dx^i=[\xi _{\text{ }}^{\overline{i}}\text{ }_{,\underline{j}%
}+(P_{jk}^i+\Gamma _{\underline{j}k}^{\overline{i}})\xi ^k]dx^j\text{ ,} \\ 
\pounds _{\partial _k}dx^i=(P_{jk}^i+\Gamma _{\underline{j}k}^{\overline{i}%
})dx^j\text{ .}
\end{array}
\label{03.12}
\end{equation}

\subsubsection{Lie derivative of covariant non-co-ordinate basic vector
fields}

In analogous way as in the case of covariant co-ordinate basic vector fields
the Lie derivatives of covariant non-co-ordinate basic vector fields can be
obtained by means of the relations 
\begin{equation}
\begin{array}{c}
\pounds _\xi [S(e^\alpha \otimes e_\beta )]=S[\pounds _\xi (e^\alpha \otimes
e_\beta )]\text{ ,} \\ 
S(e^\alpha \otimes e_\beta )=f^\alpha \text{ }_\beta \text{ ,} \\ 
\pounds _\xi [S(e^\alpha \otimes e_\beta )]=\xi ^\gamma .e_\gamma f^\alpha 
\text{ }_\beta \text{ ,} \\ 
S[\pounds _\xi (e^\alpha \otimes e_\beta )]=S(\pounds _\xi e^\alpha \otimes
e_\beta )+S(e^\alpha \otimes \pounds _\xi e_\beta )\text{ ,} \\ 
\pounds _\xi e_\beta =-(e_\beta \xi ^\gamma +C_{\beta \delta }\text{ }%
^\gamma \xi ^\delta )e_\gamma =-\xi ^\gamma \text{ }_{//\beta }.e_\gamma 
\text{ ,} \\ 
\xi ^\gamma \text{ }_{//\beta }=e_\beta \xi ^\gamma +C_{\beta \delta }\text{ 
}^\gamma \xi ^\delta \text{ ,} \\ 
\pounds _\xi e^\alpha =k^\alpha \text{ }_\gamma (\xi ).e^\gamma \text{ ,} \\ 
k^\alpha \text{ }_\gamma (\xi ).f^\gamma \text{ }_\beta =\xi ^\gamma \text{ }%
_{//\beta }.f^\alpha \text{ }_\gamma +\xi ^\gamma .e_\gamma f^\alpha \text{ }%
_\beta \text{ ,} \\ 
f^\alpha \text{ }_\gamma .f_\beta \text{ }^\gamma =g_\beta ^\alpha \text{ , }%
f^\gamma \text{ }_\beta .f_\gamma \text{ }^\alpha =g_\beta ^\alpha \text{ ,}
\\ 
e_\gamma f^\alpha \text{ }_\delta =P_{\sigma \gamma }^\alpha .f^\sigma \text{
}_\delta +\Gamma _{\delta \gamma }^\sigma .f^\alpha \text{ }_\sigma \text{ ,}
\\ 
k^\alpha \text{ }_\beta (\xi )=f^\alpha \text{ }_\gamma .\xi ^\gamma \text{ }%
_{//\delta }.f_\beta \text{ }^\delta +(P_{\beta \gamma }^\alpha +f_\beta 
\text{ }^\delta .\Gamma _{\delta \gamma }^\sigma .f^\alpha \text{ }_\sigma
)\xi ^\gamma \text{ ,}
\end{array}
\label{03.13}
\end{equation}

\begin{equation}
\begin{array}{c}
\pounds _{e_\gamma }e^\alpha =k^\alpha \text{ }_\beta (e_\gamma )e^\beta
=k^\alpha \text{ }_{\beta \gamma }.e^\beta \text{ ,} \\ 
k^\alpha \text{ }_\beta (e_\gamma )=k^\alpha \text{ }_{\beta \gamma
}=P_{\beta \gamma }^\alpha +f_\beta \text{ }^\delta (\Gamma _{\delta \gamma
}^\sigma +C_{\delta \gamma }\text{ }^\sigma )f^\alpha \text{ }_\sigma \text{
,}
\end{array}
\label{03.14}
\end{equation}

\noindent in the form 
\begin{equation}
\begin{array}{c}
\pounds _\xi e^\alpha =[\xi ^{\overline{\alpha }}\text{ }_{//\underline{%
\beta }}+(P_{\beta \gamma }^\alpha +\Gamma _{\underline{\beta }\gamma }^{%
\overline{\alpha }})\xi ^\gamma ]e^\beta = \\ 
=[e_{\underline{\beta }}\xi ^{\overline{\alpha }}+(P_{\beta \gamma }^\alpha
+\Gamma _{\underline{\beta }\gamma }^{\overline{\alpha }}+C_{\underline{%
\beta }\gamma }\text{ }^{\overline{\alpha }})\xi ^\gamma ]e^\beta \text{ ,}
\end{array}
\label{03.15}
\end{equation}
\begin{equation}
\pounds _{e_\gamma }e^\alpha =(P_{\beta \gamma }^\alpha +\Gamma _{\underline{%
\beta }\gamma }^{\overline{\alpha }}+C_{\underline{\beta }\gamma }\text{ }^{%
\overline{\alpha }})e^\beta \text{ ,}  \label{03.16}
\end{equation}

\noindent where 
\begin{equation}
\begin{array}{c}
\xi ^{\overline{\alpha }}\text{ }_{//\underline{\beta }}=f^\alpha \text{ }%
_\gamma .\xi ^\gamma \text{ }_{//\delta }.f_\beta \text{ }^\delta =f^\alpha 
\text{ }_\gamma (e_\delta \xi ^\gamma )f_\beta \text{ }^\delta +f^\alpha 
\text{ }_\gamma .C_{\delta \sigma }\text{ }^\gamma .f_\beta \text{ }^\delta
.\xi ^\sigma = \\ 
=e_{\underline{\beta }}\xi ^{\overline{\alpha }}+C_{\underline{\beta }\sigma
}\text{ }^{\overline{\alpha }}.\xi ^\sigma \text{ ,} \\ 
e_{\underline{\beta }}\xi ^{\overline{\alpha }}=f^\alpha \text{ }_\gamma
(e_\delta \xi ^\gamma )f_\beta \text{ }^\delta \text{ , }C_{\underline{\beta 
}\sigma }\text{ }^{\overline{\alpha }}=f^\alpha \text{ }_\gamma .C_{\delta
\sigma }\text{ }^\gamma .f_\beta \text{ }^\delta \text{ ,} \\ 
\Gamma _{\underline{\beta }\gamma }^{\overline{\alpha }}=f_\beta \text{ }%
^\delta .\Gamma _{\delta \gamma }^\sigma .f^\alpha \text{ }_\sigma \text{ .}
\end{array}
\label{03.17}
\end{equation}

\subsection{Classification of linear transports with respect to the
connections between contravariant and covariant affine connection}

By means of the Lie derivatives of covariant basis vector fields a
classification can be proposed for the connections between the components $%
\Gamma _{jk}^i$ ($\Gamma _{\beta \gamma }^\alpha $) of the contravariant
affine connection $\Gamma $ and the components $P_{jk}^i$ ($P_{\beta \gamma
}^\alpha $) of the covariant affine connection $P$. On this basis, linear
transports (induced by the covariant differential operator or by
connections) and draggings along (induced by the Lie-differential operator)
can be considered as connected each other through commutation relations of
both the operators with the contraction operator.

\begin{center}
$
\begin{array}{ll}
Transport\text{ condition} & Type \text{ of dragging along and transports}
\\ 
\begin{array}{c}
P_{\beta \gamma }^\alpha +\Gamma _{\underline{\beta }\gamma }^{\overline{%
\alpha }}+C_{\underline{\beta }\gamma }\text{ }^{\overline{\alpha }}=%
\overline{F}_{\beta \gamma }^\alpha \text{ ,} \\ 
P_{jk}^i+\Gamma _{\underline{j}k}^{\overline{i}}=\overline{F}_{jk}^i\text{ .}
\end{array}
& 
\begin{array}{c}
\pounds _{\epsilon _\gamma }e^\alpha = \overline{F}_{\beta \gamma }^\alpha
.e^\beta \text{ ,} \\ 
\pounds _{\partial _k}dx^i= \overline{F}_{jk}^i.dx^j\text{ .} \\ 
\text{Transport with arbitrary dragging along}
\end{array}
\\ 
\begin{array}{c}
P_{\beta \gamma }^\alpha +\Gamma _{\underline{\beta }\gamma }^{\overline{%
\alpha }}=\overline{A}_\gamma .g_\beta ^\alpha \text{ ,} \\ 
P_{jk}^i+\Gamma _{\underline{j}k}^{\overline{i}}=\overline{A}_k.g_j^i\text{ .%
}
\end{array}
& 
\begin{array}{c}
\pounds _{e_\gamma }e^\alpha = \overline{A}_\gamma .e^\alpha +C_{\underline{%
\beta }\gamma }\text{ }^{\overline{\alpha }}.e^\beta \text{ ,} \\ 
\pounds _{\partial _k}dx^i= \overline{A}_k.dx^i\text{ .} \\ 
\text{Transport with co-linear dragging along}
\end{array}
\\ 
\begin{array}{c}
P_{\beta \gamma }^\alpha +\Gamma _{\underline{\beta }\gamma }^{\overline{%
\alpha }}=0\text{ ,} \\ 
P_{jk}^i+\Gamma _{\underline{j}k}^{\overline{i}}=0\text{ .}
\end{array}
& 
\begin{array}{c}
\pounds _{e_\gamma }e^\alpha =C_{\underline{\beta }\gamma }\text{ }^{%
\overline{\alpha }}.e^\beta \text{ ,} \\ 
\pounds _{\partial _k}dx^i=0 \text{ .} \\ 
\text{Transport with invariant dragging along}
\end{array}
\end{array}
$
\end{center}

The classification of the connections on the basis of different transport
conditions is analogous to the classification, proposed by Schouten and
considered by (Schmutzer 1968).

\subsection{Lie derivatives of covariant vector fields}

The action of the Lie-differential operator on covariant vector and tensor
fields is determined by its action on covariant basic vector fields and on
the functions over $M$.

In co-ordinate basis the Lie derivative of covariant vector field $p$ along
a contravariant vector field $\xi $ can be written in the forms 
\begin{equation}
\begin{array}{c}
\pounds _\xi p=\pounds _\xi (p_idx^i)=(\pounds _\xi p_i)dx^i= \\ 
=[p_{i,k}.\xi ^k+p_j.\xi ^{\overline{j}}\text{ }_{,\underline{i}%
}+p_j(P_{ik}^j+\Gamma _{\overline{i}k}^{\overline{j}})\xi ^k]dx^i= \\ 
=[p_{i;k}\xi ^k+\xi ^{\overline{k}}\text{ }_{;\underline{i}}.p_k+T_{k%
\underline{i}}^{\overline{j}}.p_j.\xi ^k]dx^i\text{ ,}
\end{array}
\label{03.18}
\end{equation}

\noindent where 
\begin{equation}
\begin{array}{c}
\xi ^{\overline{j}}\text{ }_{;\underline{i}}=f^j\text{ }_k.\xi ^k\text{ }%
_{;l}.f_i\text{ }^l\text{ , }T_{k\underline{i}}^{\overline{j}}=f^j\text{ }%
_l.T_{km}^l.f_i\text{ }^m\text{ ,} \\ 
T_{ki}^j=\Gamma _{ik}^j-\Gamma _{ki}^j\text{ ,} \\ 
\text{(in co-ordinate basis).}
\end{array}
\label{03.19}
\end{equation}

In non-co-ordinate basis the Lie derivative $\pounds _\xi p$ has the forms 
\begin{equation}
\begin{array}{c}
\pounds _\xi p=\pounds _\xi (p_\alpha e^\alpha )=(\pounds _\xi p_\alpha
)e^\alpha = \\ 
\{(e_\gamma p_\alpha +P_{\alpha \gamma }^\beta p_\beta )\xi ^\gamma +p_\beta
[e_{\underline{\alpha }}\xi ^{\overline{\beta }}+(\Gamma _{\underline{\alpha 
}\gamma }^{\overline{\beta }}+C_{\underline{\alpha }\gamma }\text{ }^{%
\overline{\beta }})\xi ^\gamma ]\}.e^\alpha = \\ 
=(p_{\alpha /\beta }\xi ^\beta +\xi ^{\overline{\beta }}\text{ }_{/%
\underline{\alpha }}.p_\beta +T_{\gamma \underline{\alpha }}^{\overline{%
\beta }}.p_\beta .\xi ^\gamma )e^\alpha \text{ ,}
\end{array}
\label{03.20}
\end{equation}

\noindent where

\begin{equation}
\begin{array}{c}
\xi ^{\overline{\beta }}\text{ }_{/\underline{\alpha }}=f^\beta \text{ }%
_\delta .\xi ^\delta \text{ }_{/\gamma }.f_\alpha \text{ }^\gamma \text{ , }%
T_{\gamma \underline{\alpha }}^{\overline{\beta }}=f_\alpha \text{ }^\delta
.T_{\gamma \delta }^\sigma .f^\beta \text{ }_\sigma \text{ ,} \\ 
T_{\beta \gamma }^\alpha =\Gamma _{\gamma \beta }^\alpha -\Gamma _{\beta
\gamma }^\alpha -C_{\beta \gamma }\text{ }^\alpha \text{ ,} \\ 
\text{(in non-co-ordinate basis).}
\end{array}
\label{03.21}
\end{equation}

The action of the Lie-differential operator on covariant tensor fields is
determined by its action on basic tensor fields.

\section{Kinematic characteristics connected with the notion relative
velocity}

\subsection{Relative velocity}

The notion \textit{relative velocity} vector field (relative velocity) $%
_{rel}v$ can be defined as the orthogonal to a non-isotropic vector field $u$
projection of the first covariant derivative (along the same non-isotropic
vector field $u$) of (another) vector field $\xi $, i.e.

\begin{equation}  \label{2.1}
\begin{array}{c}
_{rel}v= \overline{g}(h_u(\nabla _u\xi ))=g^{ij}h_{\overline{j}\overline{k}%
}\xi ^k \text{ }_{;l}u^l.e_i= \\ 
=g^{ij}.f^m \text{ }_j.f^n\text{ }_k.h_{mn}.\xi ^k\text{ }_{;l}.u^l.e_i\text{
,} \\ 
e_i=\partial _i\text{ (in co-ordinate basis),}
\end{array}
\end{equation}

\noindent where (the indices in co-ordinate and in non-co-ordinate basis are
written in both cases as Latin indices instead as Latin and Greek indices)

\begin{equation}  \label{2.2}
h_u=g-\frac 1e.g(u)\otimes g(u)\text{ },\text{ }
\end{equation}
\[
h_u=h_{ij}.e^i.e^j\text{, }\overline{g}=g^{ij}.e_i.e_j,\text{ ,} 
\]

\begin{equation}  \label{2.3}
\begin{array}{c}
\nabla _u\xi =\xi ^i \text{ }_{;j}u^j.e_i\text{ } \\ 
\text{ }\xi ^i\text{ }_{;j}=e_j\xi ^i+\Gamma _{kj}^i\xi ^k \\ 
\text{ }\Gamma _{kj}^i\neq \Gamma _{jk}^i\text{ },
\end{array}
\end{equation}

\begin{equation}  \label{2.4}
\begin{array}{c}
g=g_{ij}.e^i.e^j, \\ 
\text{ }g_{ij}=g_{ji}\text{ }, \\ 
\text{ }e^i.e^j=\frac 12(e^i\otimes e^j+e^j\otimes e^i)\text{ },
\end{array}
\end{equation}

\begin{equation}  \label{2.5}
\begin{array}{c}
e=g(u,u)=g_{\overline{i}\overline{j}}u^iu^j=u_{\overline{i}}u^i\neq 0 \\ 
\text{ }g(u)=g_{i\overline{k}}u^k=u_i=g_{ik}.u^{\overline{k}}\text{ , }u^{%
\overline{k}}=f^k\text{ }_l.u^l\text{ ,} \\ 
\text{ }e_i.e_j=\frac 12(e_i\otimes e_j+e_j\otimes e_i)\text{ }, \\ 
g_{\overline{i}\overline{j}}=f^k\text{ }_i.f^l\text{ }_j.g_{kl}\text{ ,} \\ 
\text{ }g[\overline{g}(p)]=p\text{ , }p\in T^{*}(M)\text{ , }\overline{g}[%
g(u)]=u\text{ ,} \\ 
g^{\overline{i}\overline{k}}g_{kj}=g_j^i\text{ , g}_{\overline{i}\overline{k}%
}g^{kj}=g_i^j\text{ , g}^{\overline{i}\overline{j}}=f^i\text{ }_k.f^j\text{ }%
_l.g^{kl}\text{ ,}
\end{array}
\end{equation}

\begin{equation}  \label{2.6}
\begin{array}{c}
h_u(\nabla _u\xi )=h_{i \overline{j}}\xi ^j\text{ }_{;k}u^k.e^i \\ 
\text{ }h_{ij}=g_{ij}-\frac 1e.u_iu_j\text{ .}
\end{array}
\end{equation}

In a co-ordinate basis 
\[
\begin{array}{c}
e_j\xi ^i=\xi ^i \text{ }_{,j}=\partial _j\xi ^i=\partial \xi ^i/\partial
x^j, \\ 
e^j=dx^j, \\ 
e_i=\partial _i=\partial /\partial x^i, \\ 
u=u^i.\partial _i,
\end{array}
\]

Every contravariant vector field $\xi $ can be written by means of its
projection along and orthogonal to $u$ in two parts - one collinear to $u$
and one - orthogonal to $u$ , i.e.

\begin{equation}  \label{2.7}
\xi =\frac le.u+h^u[g(\xi )]\text{ }=\frac le.u+\overline{g}[h_u(\xi )]\text{
,}
\end{equation}

\noindent where

\begin{equation}  \label{2.8}
\begin{array}{c}
l=g(\xi ,u) \\ 
\text{ }h^u=\overline{g}-\frac 1e.u\otimes u \\ 
\text{ }\xi =\xi ^i.\partial _i=\xi ^k.e_k \\ 
\text{ }h^u=h^{ij}e_i.e_j\text{ ,}
\end{array}
\end{equation}

\begin{equation}  \label{2.9}
\begin{array}{c}
\overline{g}(h_u)\overline{g}=h^u \\ 
\text{ }h_u(\overline{g})(g)=h_u \\ 
\text{ }h^u(g)(\overline{g})=h^u \\ 
\text{ }g(h^u)g=h_u\text{ .}
\end{array}
\end{equation}

Therefore, $\nabla _u\xi $ can be written in the form

\begin{equation}  \label{2.10}
\begin{array}{c}
\nabla _u\xi = \frac{\overline{l}}e.u+\overline{g}[h_u(\nabla _u\xi )]=\frac{%
\overline{l}}e.u+_{rel}v \\ 
\text{ }\overline{l}=g(\nabla _u\xi ,u)
\end{array}
\end{equation}

\noindent and the connection between $\nabla _u\xi $ and $_{rel}v$ is
obvious. Using the relation (Yano 1957) between the Lie derivative $\pounds
_\xi u$ and the covariant derivative $\nabla _\xi u$

\begin{equation}  \label{2.11}
\begin{array}{c}
\pounds _\xi u=\nabla _\xi u-\nabla _u\xi -T(\xi ,u) \\ 
\text{ }T(\xi ,u)=T_{ij}^k\xi ^iu^j.e_k\text{ ,}
\end{array}
\end{equation}

\[
\begin{array}{c}
T_{ij}^k=-T_{ji}^k=\Gamma _{ji}^k-\Gamma _{ij}^k-C_{ij}^k \\ 
\text{ (in a non-co-ordinate basis }\{e_k\}\text{) , }
\end{array}
\]

\[
[e_i,e_j]=\pounds _{e_i}e_j\text{ }=C_{ij}^k.e_k\text{ ,} 
\]

\[
\begin{array}{c}
T_{ij}^k=\Gamma _{ji}^k-\Gamma _{ij}^k \\ 
\text{ (in a co-ordinate basis }\{\partial _k\}\text{ ) ,}
\end{array}
\]

\noindent one can write $\nabla _u\xi $ in the form

\begin{equation}  \label{2.12}
\nabla _u\xi =(k)g(\xi )-\pounds _\xi u\text{ }=k[g(\xi )]-\pounds _\xi u%
\text{,}
\end{equation}

\noindent or taking into account the above expression for $\xi $ - in the
form

\[
\nabla _u\xi =k[h_u(\xi )]+\frac le.a-\pounds _\xi u\text{ ,} 
\]

\noindent where

\begin{equation}
\begin{array}{c}
k[g(\xi )]=\nabla _\xi u-T(\xi ,u) \\ 
\text{ }k=(u^i\text{ }_{;l}-T_{lk}^iu^k)g^{lj}.e_i\otimes e_j\text{ ,}
\end{array}
\label{2.13}
\end{equation}

\begin{equation}  \label{2.14}
\begin{array}{c}
k[g(u)]=k(g)u=k^{ij}g_{\overline{j}\overline{k}}u^k.e_i \\ 
=a=\nabla _uu=u^i\text{ }_{;j}u^j.e_i \text{ .}
\end{array}
\end{equation}

For $h_u(\nabla _u\xi )$ it follows

\begin{equation}  \label{2.15}
h_u(\nabla _u\xi )=h_u(\frac le.a-\pounds _\xi u)+h_u(k)h_u(\xi ) \text{ ,}
\end{equation}

\noindent where 
\[
\begin{array}{c}
h_u(k)h_u(\xi )=h_{i\overline{k}}k^{kl}h_{\overline{l}\overline{j}}\xi
^j.e^i, \\ 
\text{ }h_u(u)=0, \\ 
u(h_u)=0, \\ 
\text{ }h_u(k)h_u(u)=0, \\ 
\text{ }(u)h_u(k)h_u=0.
\end{array}
\]

If we introduce the abbreviation

\begin{equation}  \label{2.16}
d=h_u(k)h_u=h_{i\overline{k}}k^{kl}h_{\overline{l}j}.e^i\otimes
e^j=d_{ij}.e^i\otimes e^j\text{ ,}
\end{equation}

\noindent the expression for $_{rel}v$ can take the form

\[
_{rel}v=\overline{g}[h_u(\nabla _u\xi )]=\overline{g}(h_u)(\frac
le.a-\pounds _\xi u)+\overline{g}[d(\xi )]= 
\]

\begin{equation}  \label{2.17}
=[g^{ik}h_{\overline{k}\overline{l}}(\frac le.a^l-\pounds _\xi u^l)+g^{ik}d_{%
\overline{k}\overline{l}}\xi ^l].e_i\text{ }=_{rel}v^i.e_i \text{ ,}
\end{equation}

\noindent or

\begin{equation}  \label{2.18}
g(_{rel}v)=h_u(\nabla _u\xi )=h_u(\frac le.a-\pounds _\xi u)+d(\xi )\text{ .}
\end{equation}

For the special case when the vector field $\xi $ is orthogonal to $u$, i.e. 
$\xi =\overline{g}[h_u(\xi )]$, and the Lie derivative of $u$ along $\xi $
is zero, i.e. $\pounds _\xi u=0$, then the relative velocity can be written
in the form 
\begin{equation}  \label{2.19}
g(_{rel}v)=d(\xi )
\end{equation}

\noindent or in the form 
\[
_{rel}v=\overline{g}[d(\xi )]\text{.} 
\]

\subsection{Deformation velocity, shear velocity, rotation velocity and
expansion velocity}

The covariant tensor field $d$ is a generalization for $\overline{(L}_n,g)$%
-spaces of the well known \textit{deformation velocity }tensor for $V_n$%
-spaces (Stephani 1977), (Kramer, Stephani, MacCallum, Herlt 1980). It is
usually represented by means of its three parts: the trace-free symmetric
part, called \textit{shear velocity }tensor (shear), the anti symmetric
part, called \textit{rotation velocity }tensor (rotation) and the trace
part, in which the trace is called \textit{expansion velocity }(expansion)%
\textit{\ }invariant.

After some more complicated as for $V_n$-spaces calculations the deformation
velocity tensor $d$ can be given in the form

\begin{equation}  \label{2.20}
\begin{array}{c}
d=h_u(k)h_u=h_u(k_s)h_u+h_u(k_a)h_u= \\ 
=\sigma +\omega +\frac 1{n-1}.\theta .h_u\text{ ,}
\end{array}
\end{equation}

\noindent where

$\sigma $ is the \textit{shear velocity} tensor (shear) , 
\begin{equation}  \label{2.21}
\begin{array}{c}
\sigma =_sE-_sP=E-P-\frac 1{n-1}. \overline{g}[E-P].h_u=\sigma _{ij}.e^i.e^j=
\\ 
=E-P-\frac 1{n-1}.(\theta _o-\theta _1).h_u,
\end{array}
\end{equation}

\begin{equation}  \label{2.22}
\begin{array}{c}
_sE=E-\frac 1{n-1}. \overline{g}[E].h_u \\ 
\text{ }\overline{g}[E]=g^{ij}.E_{\overline{i} \overline{j}}=g^{\overline{i}%
\overline{j}}.E_{ij}=\theta _o\text{ ,}
\end{array}
\end{equation}

\begin{equation}  \label{2.23}
\begin{array}{c}
E=h_u(\epsilon )h_u \\ 
\text{ }k_s=\epsilon -m \\ 
\text{ }\epsilon =\frac 12(u_{\text{ };l}^ig^{lj}+u_{\text{ }%
;l}^j.g^{li}).e_i.e_j\text{ ,}
\end{array}
\end{equation}

\begin{equation}  \label{2.24}
m=\frac 12(T_{lk}^iu^kg^{lj}+T_{lk}^ju^kg^{li})e_i.e_j\text{ .}
\end{equation}

$_sE$ is the \textit{torsion-free shear velocity }tensor, $_sP$ is the 
\textit{shear velocity} tensor \textit{induced by the torsion},

\textit{
\begin{equation}  \label{2.25}
\begin{array}{c}
_sP=P-\frac 1{n-1}. \overline{g}[P].h_u \\ 
\text{ }\overline{g}[P]=g^{kl}P_{\overline{k} \overline{l}}\text{ }=g^{%
\overline{k}\overline{l}}.P_{kl}=\theta _1\text{,}
\end{array}
\end{equation}
}

\textit{
\begin{equation}  \label{2.26}
\begin{array}{c}
P=h_u(m)h_u \\ 
\text{ }\theta _1=T_{kl}^ku^l \\ 
\text{ }\theta _o=u^n\text{ }_{;n}-\frac 1{2e}(e_{,k}u^k-g_{kl;m}u^mu^{%
\overline{k}}u^{\overline{l}})\text{ ,}
\end{array}
\end{equation}
}

\textit{
\begin{equation}  \label{2.27}
\begin{array}{c}
e_{,k}=e_ke \\ 
\text{ }\theta =\theta _o-\theta _1\text{ , }
\end{array}
\end{equation}
}

$\theta $ is the \textit{expansion velocity, $\theta _o$} is the \textit{%
torsion-free expansion velocity,} $\theta _1$ is the \textit{expansion
velocity induced by the torsion,}

$\omega $ is the\textit{\ rotation velocity }tensor (rotation velocity),

\begin{equation}  \label{2.28}
\omega =h_u(k_a)h_u=h_u(s)h_u-h_u(q)h_u=S-Q\text{ ,}
\end{equation}

\begin{equation}  \label{2.29}
\begin{array}{c}
s=\frac 12(u^k \text{ }_{;m}g^{ml}-u^l\text{ }_{;m}g^{mk}).e_k\wedge e_l \\ 
\text{ }e_k\wedge e_l=\frac 12(e_k\otimes e_l-e_l\otimes e_k)\text{ , }
\end{array}
\end{equation}

\begin{equation}  \label{2.30}
\begin{array}{c}
q=\frac 12(T_{mn}^kg^{ml}-T_{mn}^lg^{mk})u^n.e_k\wedge e_l \\ 
\text{ }S=h_u(s)h_u\text{ , }Q=h_u(q)h_u\text{ ,}
\end{array}
\end{equation}

$S$ is the \textit{torsion-free rotation velocity} tensor, $Q$ is the\textit{%
\ rotation velocity }tensor \textit{induced by the torsion.}

By means of the expressions for $\sigma $, $\omega $ and $\theta $ the
deformation velocity tensor can be written in two parts

\begin{equation}  \label{2.31}
\begin{array}{c}
d=d_o+d_1 \\ 
\text{ }d_o=_sE+S+\frac 1{n-1}.\theta _o.h_u \\ 
\text{ }d_1=_sP+Q+\frac 1{n-1}.\theta _1.h_u\text{ ,}
\end{array}
\end{equation}

\noindent where $d_o$ is the \textit{torsion-free deformation velocity}
tensor and $d_1$ is the \textit{deformation velocity }tensor \textit{induced
by the torsion. }For the case of $V_n$-spaces $d_1=0$ ($_sP=0$ , $Q=0$ , $%
\theta _1=0$).

The shear velocity tensor $\sigma $ and the expansion velocity $\theta $ can
be written also in the form

\begin{equation}  \label{2.32}
\sigma =\frac 12\{h_u(\nabla _u\overline{g}-\pounds _u\overline{g})h_u-\frac
1{n-1}(h_u[\nabla _u\overline{g}-\pounds _u\overline{g}]).h_u\} \text{ }=
\end{equation}

\begin{equation}  \label{2.33}
=\frac 12\{h_{i\overline{k}}(g^{kl}\text{ }_{;m}u^m-\pounds _ug^{kl})h_{%
\overline{l}j}-\frac 1{n-1}.h_{\overline{k}\overline{l}}(g^{kl} \text{ }%
_{;m}u^m-\pounds _ug^{kl}).h_{ij}\}.e^i.e^j\text{ .}
\end{equation}
\[
\begin{array}{c}
\theta =\frac 12.h_u[\nabla _u \overline{g}-\pounds _u\overline{g}]=\frac
12[\nabla _{\overline{g}}u+T(u, \overline{g})]= \\ 
=\frac 12h_{\overline{i}\overline{j}}(g^{ij}\text{ }_{;k}u^k-\pounds
_ug^{ij})\text{ .}
\end{array}
\]

The main result of the above considerations can be summarized in the
following proposition:

\textbf{Proposition 2.}The covariant vector field $g(_{rel}v)=h_u(\nabla
_u\xi ) $ can be written in the forms:

\[
h_u(\nabla _u\xi )=h_u(\frac le.a-\pounds _\xi u)+d(\xi )= 
\]

\[
=h_u(\frac le.a-\pounds _\xi u)+\sigma (\xi )+\omega (\xi )+\frac
1{n-1}.\theta .h_u(\xi )\text{ .} 
\]

The physical interpretation of the velocity tensors $d,$ $\sigma ,$ $\omega $
and of the invariant $\theta $ for the case of $V_4$-spaces (Synge 1960),
(Ehlers 1961), (Kramer, Stephani, MacCallum, Herlt 1980) can be extended
also for $(\overline{L}_4,g)$-spaces. In this case the torsion play an
equivalent role as the covariant derivative in the velocity tensors. The
individual designation, connected with the physical interpretation of these
kinematic characteristics, is given in the Appendix A - Table $1$. It is
easy to see that the existence of some kinematic characteristics ($_sP$, $Q$%
, $\theta _1$) depends on the existence of the torsion tensor field. They
vanish if it is equal to zero (e.g. in $V_n$-spaces).

\section{Kinematic characteristics connected with the notion relative
acceleration}

\subsection{Relative acceleration}

The notion \textit{relative acceleration} vector field (relative
acceleration) $_{rel}a$ can be defined (in analogous way as $_{rel}v$) as
the orthogonal to a non-isotropic vector field $u$ ($g(u,u)=e\neq 0$)
projection of the second covariant derivative (along the same non-isotropic
vector field $u$) of (another) vector field $\xi $, i.e. 
\begin{equation}  \label{3.1}
_{rel}a=\overline{g}(h_u(\nabla _u\nabla _u\xi ))=g^{ij}h_{\overline{j}%
\overline{k}}(\xi ^k\text{ }_{;l}u^l)_{;m}u^me_i\text{ .}
\end{equation}

$\nabla _u\nabla _u\xi $ $=(\xi ^i$ $_{;l}u^l)_{;m}u^me_i$ is the second
covariant derivative of a vector field $\xi $ along the vector field $u$. It
is an essential part of all types of deviation equations in $V_n$- and ($%
L_n,g$)-spaces (Manoff 1979,1984), (Iliev, Manoff 1983).

If we take into account the expression for $\nabla _u\xi $%
\[
\nabla _u\xi =k[g(\xi )]-\pounds _\xi u\text{,} 
\]

\noindent and differentiate covariant along $u$, then we obtain 
\[
\nabla _u\nabla _u\xi =\{\nabla _u[(k)g]\}(\xi )+(k)(g)(\nabla _u\xi
)-\nabla _u(\pounds _\xi u) 
\]

By means of the relations 
\[
k(g)\overline{g}=k\text{ ,} 
\]
\[
\nabla _u[k(g)]=(\nabla _uk)(g)+k(\nabla _ug)\text{ ,} 
\]
\[
\{\nabla _u[k(g)]\}\overline{g}=\nabla _uk+k(\nabla _ug)\overline{g}\text{ ,}
\]

\noindent $\nabla _u\nabla _u\xi $ can be written in the form 
\begin{equation}
\nabla _u\nabla _u\xi =\frac le.H(u)+B(h_u)\xi -k(g)\pounds _\xi u-\nabla
_u(\pounds _\xi u)  \label{3.2}
\end{equation}

(compare with $\nabla _u\xi =\frac le.a+k(h_u)\xi -\pounds _\xi u$),

\noindent where 
\[
H=B(g)=(\nabla _uk)(g)+k(\nabla _ug)+k(g)k(g)\text{ ,} 
\]
\[
B=\nabla _uk+k(g)k+k(\nabla _ug)\overline{g}=\nabla _uk+k(g)k-k(g)(\nabla _u%
\overline{g})\text{ .} 
\]

The orthogonal to $u$ covariant projection of $\nabla _u\nabla _u\xi $ will
have therefore the form 
\begin{equation}  \label{3.3}
h_u(\nabla _u\nabla _u\xi )=h_u[\frac leH(u)-k(g)\pounds _\xi u-\nabla
_u\pounds _\xi u]+[h_u(B)h_u](\xi )\text{ .}
\end{equation}

In the special case, when $g(u,\xi )=l=0$ and $\pounds _\xi u=0$ , the above
expression has the simple form 
\begin{equation}  \label{3.4}
h_u(\nabla _u\nabla _u\xi )=[h_u(B)h_u](\xi )=A(\xi )\text{ ,}
\end{equation}

(compare with $h_u(\nabla _u\xi )=[h_u(k)h_u](\xi )=d(\xi )$).

The explicit form of $H(u)$ follows from the explicit form of $H$ and its
action on the vector field $u$%
\begin{equation}  \label{3.5}
H(u)=(\nabla _uk)[g(u)]+k(\nabla _ug)(u)+k(g)(a)=\nabla _u[k(g)(u)]=\nabla
_ua\text{ .}
\end{equation}

Now $h_u[\nabla _u\nabla _u\xi ]$ can be written in the form 
\begin{equation}  \label{3.6}
h_u(\nabla _u\nabla _u\xi )=h_u[\frac le.\nabla _ua-k(g)(\pounds _\xi
u)-\nabla _u(\pounds _\xi u)]+A(\xi )
\end{equation}

(compare $h_u(\nabla _u\xi )=h_u(\frac le.a-\pounds _\xi u)+d(\xi )$).

The explicit form of $A=h_u(B)h_u$ can be found in analogous way as the
explicit form for $d=h_u(k)h_u$ in the expression for $_{rel}v$.

\subsection{Deformation acceleration, shear acceleration, rotation
acceleration and expansion acceleration}

The covariant tensor $A$, named \textit{deformation acceleration} tensor can
be represented as a sum, containing three terms: a trace-free symmetric
term, an anti symmetric term and a trace term 
\begin{equation}  \label{3.7}
A=_sD+W+\frac 1{n-1}.U.h_u
\end{equation}

\noindent where 
\begin{equation}
D=h_u(_sB)h_u  \label{3.8}
\end{equation}
\begin{equation}
W=h_u(_aB)h_u  \label{3.9}
\end{equation}
\begin{equation}
U=\overline{g}[_sA]=\overline{g}[D]  \label{3.10}
\end{equation}
\begin{equation}
_sB=\frac 12(B^{ij}+B^{ji})e_i.e_j\text{ , }_aB=\frac
12(B^{ij}-B^{ji})e_i\wedge e_j\text{,}  \label{3.11}
\end{equation}
\begin{equation}
_sA=\frac 12(A_{ij}+A_{ji})e^i.e^j\text{ },  \label{3.12}
\end{equation}
\begin{equation}
_sD=D-\frac 1{n-1}.\overline{g}[D].h_u=D-\frac 1{n-1}.U.h_u\text{ .}
\label{3.13}
\end{equation}

$_sD$ is the \textit{shear acceleration} tensor (shear acceleration), $W$ is
the \textit{rotation acceleration} tensor (rotation acceleration) and $U$ is
the \textit{expansion acceleration }invariant (expansion acceleration).
Furthermore, every one of these quantities can be divided into three parts:
torsion- and curvature-free acceleration, acceleration induced by torsion
and acceleration induced by curvature.

Let us now consider the representation of every acceleration quantity in its
essential parts, connected with its physical interpretation.

The deformation acceleration tensor $A$ can be written in the following
forms 
\begin{equation}  \label{3.14}
A=_sD+W+\frac 1{n-1}.U.h_u=A_0+G=_FA_0-_TA_0+G\text{ ,}
\end{equation}
\begin{equation}  \label{3.15}
A=_sD_0+W_0+\frac 1{n-1}.U_0.h_u+_sM+N+\frac 1{n-1}.I.h_u\text{ ,}
\end{equation}
\begin{equation}  \label{3.16}
\begin{array}{c}
A=_{sF}D_0+_FW_0+\frac 1{n-1}._FU_0.h_u- \\ 
-(_{sT}D_0+_TW_0+\frac 1{n-1}._TU_{0.}h_u)+ \\ 
+_sM+N+\frac 1{n-1}.I.h_u\text{ ,}
\end{array}
\end{equation}

\noindent where 
\begin{equation}
A_0=_FA_0-_TA_0=_sD_0+W_0+\frac 1{n-1}.U_0.h_u\text{ ,}  \label{3.17}
\end{equation}
\begin{equation}
_FA_0=_{sF}D_0+_FW_0+\frac 1{n-1}._FU_0.h_u\text{ ,}  \label{3.18}
\end{equation}
\begin{equation}
_FA_0(\xi )=h_u(\nabla _{\xi _{\perp }}a)\text{ , }\xi _{\perp }=\overline{g%
}[h_u(\xi )]\text{ ,}  \label{3.19}
\end{equation}
\begin{equation}
_TA_0=_{sT}D_0+_TW_0+\frac 1{n-1}._TU_0.h_u\text{ ,}  \label{3.20}
\end{equation}
\begin{equation}
G=_sM+N+\frac 1{n-1}.I.h_u=h_u(K)h_u\text{ ,}  \label{3.21}
\end{equation}
\begin{equation}
h_u([R(u,\xi )]u)=h_u(K)h_u(\xi )\text{ for }\forall \text{ }\xi \in T(M)%
\text{ , }  \label{3.22}
\end{equation}
\begin{equation}
\lbrack R(u,\xi )]u=\nabla _u\nabla _\xi u-\nabla _\xi \nabla _uu-\nabla
_{\pounds _u\xi }u\text{ ,}  \label{3.23}
\end{equation}
\begin{equation}
K=K^{kl}e_k\otimes e_l\text{ , }K^{kl}=R^k\text{ }_{mnr}g^{rl}u^mu^n\text{ ,}
\label{2.34}
\end{equation}

$R^k$ $_{mnr}$ are the components of the contravariant Riemannian curvature
tensor, 
\begin{equation}
K_a=K_a^{kl}.e_k\wedge e_l\text{, }K_a^{kl}=\frac 12(K^{kl}-K^{lk})\text{ , }%
K_s=K_s^{kl}e_k.e_l\text{ , }K_s^{kl}=\frac 12(K^{kl}+K^{lk})\text{ ,}
\label{2.35}
\end{equation}
\begin{equation}
_sD=_sD_0+_sM\text{ , }W=W_0+N=_FW_0-_TW_0+N\text{ ,}  \label{2.36}
\end{equation}
\begin{equation}
U=U_0+I=_FU_0-_TU_0+I\text{ ,}  \label{2.37}
\end{equation}
\begin{equation}
_sM=M-\frac 1{n-1}.I.h_u\text{ , }M=h_u(K_s)h_u\text{ , }I=\overline{g}[M%
]=g^{\overline{i}\overline{j}}M_{ij}\text{ ,}  \label{2.38}
\end{equation}
\begin{equation}
N=h_u(K_a)h_u\text{ ,}  \label{2.39}
\end{equation}
\begin{equation}
_sD_0=_{sF}D_0-_{sT}D_0=_FD_0-\frac 1{n-1}._FU_0.h_u-(_TD_0-\frac
1{n-1}._TU_0.h_u)\text{ ,}  \label{2.40}
\end{equation}
\begin{equation}
_sD_0=_{sF}D_0-_TD_0-\frac 1{n-1}(_FU_0-_TU_0)h_u\text{ ,}  \label{2.41}
\end{equation}
\begin{equation}
_sD_0=D_0-\frac 1{n-1}.U_0.h_u\text{ ,}  \label{2.42}
\end{equation}
\begin{equation}
_{sF}D_0=_FD_0-\frac 1{n-1}._FU_0.h_u\text{ , }_FD_0=h_u(b_s)h_u\text{ ,}
\label{2.43}
\end{equation}
\begin{equation}
b=b_s+b_a\text{ , }b=b^{kl}e_k\otimes e_l\text{ , }b^{kl}=a^k\text{ }%
_{;n}g^{nl}\text{ ,}  \label{2.44}
\end{equation}
\begin{equation}
a^k=u^k\text{ }_{;m}u^m\text{ , }b_s=b_s^{kl}e_k.e_l\text{ , }b_s^{kl}=\frac
12(b^{kl}+b^{lk})\text{ ,}  \label{2.45}
\end{equation}
\begin{equation}
b_a=b_a^{kl}e_k\wedge e_l\text{ , }b_a^{kl}=\frac 12(b^{kl}-b^{lk})\text{ ,}
\label{2.46}
\end{equation}
\begin{equation}
_FU_0=\overline{g}[_FD_0]=g[b]-\frac 1e.g(u,\nabla _ua)\text{ , }g[b]=g_{%
\overline{k}\overline{l}}b^{kl}\text{ ,}  \label{2.47}
\end{equation}
\begin{equation}
_{sT}D_0=_TD_0-\frac 1{n-1}._TU_0.h_u=_{sF}D_0-_sD_0\text{ , }_TD_0=_FD_0-D_0%
\text{ ,}  \label{2.48}
\end{equation}
\begin{equation}
U_0=\overline{g}[D_0]=_FU_0-_TU_0\text{ , }_TU_0=\overline{g}[_TD_0]\text{ ,}
\label{2.49}
\end{equation}
\begin{equation}
_FW_0=h_u(b_a)h_u\text{ , }_TW_0=_FW_0-W_0\text{ .}  \label{2.50}
\end{equation}

Under the conditions $\pounds _\xi u=0$ , $\xi =\xi _{\perp }=\overline{g}%
(h_u(\xi ))$ , ($l=0$), the expression for $h_u(\nabla _u\nabla _u\xi )$ can
be written in the forms 
\begin{equation}
h_u(\nabla _u\nabla _u\xi _{\perp })=A(\xi _{\perp })=A_0(\xi _{\perp
})+G(\xi _{\perp })\text{ ,}  \label{2.51}
\end{equation}
\begin{equation}
h_u(\nabla _u\nabla _u\xi _{\perp })=_FA_0(\xi _{\perp })-_TA_0(\xi _{\perp
})+G(\xi _{\perp })\text{ ,}  \label{2.52}
\end{equation}
\[
h_u(\nabla _u\nabla _u\xi _{\perp })=(_{sF}D_0+_FW_0+\frac
1{n-1}._FU_0.g)(\xi _{\perp })- 
\]
\begin{equation}
-(_{sT}D_0+_TW_0+\frac 1{n-1}._TU_0.g)(\xi _{\perp })+(_sM+N+\frac
1{n-1}.I.g)(\xi _{\perp })\text{ ,}  \label{2.53}
\end{equation}

\noindent which enable one to find a physical interpretation of the
quantities $_sD$,$W$,$U$ and of the contained in their structure quantities$%
_{sF}D_0$, $_FW_0$, $_FU_0$, $_{sT}D_0$, $_TW_0$, $_TU_0$, $_sM$, $N$, $I$.
The individual designation, connected with their physical interpretation, is
given in the Appendix A - Table $1$. The expressions of these quantities in
terms of the kinematic characteristics of the relative velocity are given in
the Appendix A.

After the above consideration the following proposition can be formulated:

\textbf{Proposition 3. }$g(_{rel}a)=h_u(\nabla _u\nabla _u\xi )$ can be
written in the form 
\[
g(_{rel}a)=h_u[\frac le.\nabla _ua-\nabla _{\pounds _\xi u}u-\nabla
_u(\pounds _\xi u)+T(\pounds _\xi u,u)]+A(\xi )\text{ ,} 
\]

\noindent where 
\begin{equation}
A(\xi )=_sD(\xi )+W(\xi )+\frac 1{n-1}.U.h_u(\xi )\text{ .}  \label{2.54}
\end{equation}

For the case of affine symmetric connection ($T(w,v)=0$ for $\forall $ $%
w,v\in T(M)$ , $T_{ij}^k=0$, $\Gamma _{ij}^k=\Gamma _{ji}^k$ ) and
Riemannian metric ($\nabla _vg=0$ for $\forall v\in T(M)$, $g_{ij;k}=0$)
kinematic characteristics are obtained in $V_n$-spaces, connected with the
notion relative velocity (Manoff 1992) and relative acceleration (Manoff
1985). For the case of affine non-symmetric connection ($T(w,v)\neq 0$ for $%
\forall $ $w,v\in T(M)$ , $\Gamma _{jk}^i\neq \Gamma _{kj}^i$) and
Riemannian metric kinematic characteristics are obtained in $U_n$-spaces
(Manoff 1985).

\section{Classification of auto-parallel vector fields on the basis of the
kinematic characteristics connected with the relative velocity and relative
acceleration}

The classification of (pseudo)Riemannian spaces $V_n$, admitting the
existence of auto-parallel (in the case of $V_n$-spaces they are geodesic)
vector fields ($\nabla _uu=a=0$) with given kinematic characteristics,
connected with the notion relative velocity, can be extended to a
classification of differentiable manifolds with contravariant and covariant
affine connection and metric, admitting auto parallel vector fields with
certain kinematic characteristics, connected with the relative velocity and
the relative acceleration. In this way the following two schemes for the
existence of special type $1$. and $2$. of vector fields can be proposed (s.
Appendix B. - Table $2$.). Different types of combinations between the
single conditions of the two schemes can also be taken under consideration.

\subsection{Special geodesic vector fields with vanishing kinematic
characteristics, induced by the curvature, in (pseudo) Riemannian spaces}

On the basis of the classification $2$. the following propositions in the
case of $V_n$-spaces can be proved:

\textbf{Proposition 4. }Non-isotropic geodesic vector fields in $V_n$-spaces
are geodesic vector fields with curvature rotation acceleration tensor $N$
equal to zero, i.e. $N=0$.

Proof: 
\begin{equation}
\begin{array}{c}
N=h_u(K_a)h_u=h_{ik}K_a^{kl}h_{lj}e^i\wedge e^j, \\ 
K_a^{kl}=\frac 12(K^{kl}-K^{lk})=\frac
12(R_{mnr}^kg^{rl}-R_{mnr}^lg^{rk})u^mu^n,
\end{array}
\label{4.1}
\end{equation}

For the case of $V_n$-space, where 
\begin{equation}
R_{kmnr}=R_{nrkm}\text{ , }R_{kmnr}=g_{kl}R^l\text{ }_{mnr}\text{ ,}
\label{4.2}
\end{equation}

\noindent the conditions 
\begin{equation}
R^k\text{ }_{mnr}g^{rl}=R^l\text{ }_{nmr}g^{rk}\text{ , }R^k\text{ }_{mn}%
\text{ }^l=R^l\text{ }_{nm}\text{ }^k\text{,}  \label{4.3}
\end{equation}

\noindent follow and therefore

\begin{equation}
K_a^{kl}=\frac 12(R^k\text{ }_{mnr}g^{rl}-R^l\text{ }_{mnr}g^{rk})u^mu^n=0%
\text{ , }  \label{4.4}
\end{equation}
\[
K_a=0\text{ , }N=0\text{ .} 
\]

\textbf{Proposition 5. }Non-isotropic geodesic vector fields in $V_n$-spaces
with equal to zero Ricci tensor ($R_{ik}=R^l$ $_{ikl}=g_m^lR^m$ $_{ikl}=0$)
are geodesic vector fields with curvature rotation acceleration $N$ and
curvature expansion acceleration $I$, both equal to zero, i.e. $N=0$, $I=0$.

Proof: $1$. From the proposition $4$. it follows that $K_a=0$ and $N=0$. 
\begin{equation}
2\text{. }I=g[K]=g_{ij}K^{ij}=g_{ij}R^i\text{ }_{mnr}g^{rj}u^mu^n=g_i^rR^i%
\text{ }_{mnr}u^mu^n=R_{mn}u^mu^n=0\text{ .}  \label{4.5}
\end{equation}

\textbf{Proposition 6. }Non-isotropic geodesic vector fields in $V_n$-spaces
with constant curvature 
\begin{equation}
\lbrack R(\xi ,\eta )]v=\frac 1{n(n-1)}.R_0[g(v,\xi )\eta -g(v,\eta )\xi ]%
\text{ , }\forall \xi ,\eta ,v\in T(M)\text{,}  \label{4.6}
\end{equation}

(in index form 
\begin{equation}
R^i\text{ }_{jkl}=\frac{R_0}{n(n-1)}(g_l^i.g_{jk}-g_k^i.g_{jl})\text{ , }%
R_0=const\text{.)}  \label{4.7}
\end{equation}

\noindent are geodesic vector fields with curvature shear acceleration and
curvature rotation acceleration, both equal to zero, i.e. $_sM=0$, $N=0$.

Proof: $1$. From the proposition $4$. it follows that $N=0$. 
\[
2\text{. }_sM=M-\frac 1{n-1}.I.h_u\text{ , }M=h_u(K_s)h_u=g(K_s)g\text{ ,} 
\]
\[
M=g(K_s)g=g_{ik}K^{kl}g_{lj}e^i.e^j=M_{ij}e^i.e^j\text{ ,} 
\]
\begin{equation}
M_{ij}=g_{ik}R^k\text{ }_{mnr}g^{rl}g_{lj}u^mu^n=R_{imnj}u^mu^n=\frac{R_0}{%
n(n-1)}.e.h_{ij}\text{ ,}  \label{4.8}
\end{equation}
\begin{equation}
M=\frac{R_0.e}{n(n-1)}.h_u\text{ , }e=g(u,u)=g_{ij}u^iu^j\text{ ,}
\label{4.9}
\end{equation}
\begin{equation}
I=g[K]=\overline{g}[M]=g^{ij}M_{ij}=\frac 1n.R_0.e\text{ , }g^{ij}h_{ij}=n-1%
\text{ ,}  \label{4.10}
\end{equation}
\begin{equation}
_sM=M-\frac 1{n-1}.I.h_u=0\text{.}  \label{4.11}
\end{equation}

The projections of the curvature tensor of the type $G=h_u(K)h_u$ (or $R^i$ $%
_{jkl}u^ju^k$) along the non-isotropic vector field $u$ acquire a natural
physical meaning as quantities, connected with the kinematic characteristics
curvature shear acceleration $_sM$, curvature rotation acceleration $N$ and
curvature expansion acceleration $I$.

The projection of the Ricci tensor ($g[K]$, or $R_{ik}u^iu^k$) and the
Raychaudhuri identity for vector fields represent an expression of the
curvature expansion acceleration, given in terms of the kinematic
characteristics of the relative velocity 
\begin{equation}
\begin{array}{c}
I=\overline{g}[M]=R_{ij}u^iu^j= \\ 
=-a^j\text{ }_{;j}+g^{\overline{i}\overline{j}}._sE_{ik}.g^{\overline{k}%
\overline{l}}.\sigma _{lj}+g^{\overline{i}\overline{j}}S_{ik}g^{\overline{k}%
\overline{l}}\omega _{lj}+\theta _0^{.}+\frac 1{n-1}.\theta _0.\theta + \\ 
+\frac 1e[a^k(e_{,k}-u_{\overline{n}}T_{km}^nu^m-g_{mn;k}u^{\overline{m}}u^{%
\overline{n}}-g_{\overline{k}\overline{m};l}u^lu^m)+ \\ 
+\frac 12(u^ke_{,k})_{;l}u^l-\frac 12(g_{mn;k}u^k)_{;l}u^lu^{\overline{m}}u^{%
\overline{n}}]- \\ 
\frac 1{e^2}[\frac 34(e_{,k}u^k)^2-(e_{,k}u^k)g_{mn;l}u^lu^{\overline{m}}u^{%
\overline{n}}+\frac 14(g_{mn;l}u^lu^{\overline{m}}u^{\overline{n}})^2]\text{
,} \\ 
\theta ^{.}=\theta _{,k}u^k
\end{array}
\label{4.12}
\end{equation}

In the case of $V_n$-spaces the kinematic characteristics, connected with
the relative velocity and the relative acceleration have the forms:

a) kinematic characteristics, connected with the relative velocity

$
\begin{array}{ccc}
d=d_0 & d_1=0 & k=k_o \\ 
\sigma =_sE & _sP=0 & m=0 \\ 
\omega =S & Q=0 & q=0 \\ 
\theta =\theta _o & \theta _1=0 & \nabla _uu=a\neq 0\text{ , }a=0
\end{array}
$

b) kinematic characteristics, connected with the relative acceleration ($%
\nabla _uu=a\neq 0$)

$
\begin{array}{ccc}
A=_FA_0+G & _TA_0=0 & N=0 \\ 
G=_sM+\frac 1{n-1}.I.h_u & _{sT}D_0=0 &  \\ 
W=_FW_0 & _TW_0=0 &  \\ 
U=_FU_0+I & _TU_0=0 & \nabla _uu=a\neq 0
\end{array}
$

c) kinematic characteristics, connected with the relative acceleration ($%
\nabla _uu=a=0$)

$
\begin{array}{c}
A=G \\ 
G=_sM+\frac 1{n-1}.I.h_u \\ 
W=0 \\ 
U=I
\end{array}
\begin{array}{cc}
_TA_0=0 & N=0 \\ 
_{sT}D_0=0 &  \\ 
_TW_0=0 &  \\ 
_TU_0=0 & \nabla _uu=a=0
\end{array}
$

On the basis of the different kinematic characteristics dynamic systems can
be classified and considered in $V_n$-spaces.

\subsection{Special vector fields over manifolds with contravariant and
covariant affine connection and metric with vanishing kinematic
characteristics induced by the curvature}

The explicit forms of the quantities $G$, $M$, $N$ and $I$, connected with
accelerations induced by curvature can be used for finding conditions for
existence of special types of contravariant vector fields with vanishing
characteristics induced by the curvature. $G$, $M$, $N$ and $I$ can be
expressed in the following forms: 
\begin{equation}
\begin{array}{c}
G=h_u(K)h_u=g(K)g-\frac 1e.g(u)\otimes [g(u)](K)g\text{ , } \\ 
K[g(u)]=0\text{ ,}
\end{array}
\label{4.13}
\end{equation}
\begin{equation}
\begin{array}{c}
M=h_u(K_s)h_u=g(K_s)g-\frac 1{2e}\{g(u)\otimes [g(u)](K)g+[g(u)](K)g\otimes
g(u)\}= \\ 
=M_{ij}.dx^i.dx^j=M_{\alpha \beta }.e^\alpha .e^\beta \text{ , }M_{ij}=M_{ji}%
\text{ ,} \\ 
M_{ij}=\frac 12[g_{i\overline{k}}.g_{\overline{l}j}+g_{j\overline{k}}.g_{%
\overline{l}i}-\frac 1e(u_i.g_{\overline{l}j}+u_j.g_{\overline{l}i})u_{%
\overline{k}}]R^k\text{ }_{mnq}u^mu^n.g^{ql}\text{ ,}
\end{array}
\label{4.14}
\end{equation}
\begin{equation}
I=\overline{g}[M]=g[K_s]=g[K]=R_{\rho \sigma }.u^\rho u^\sigma =R_{kl}.u^ku^l%
\text{ ,}  \label{4.15}
\end{equation}
\begin{equation}
\begin{array}{c}
N=h_u(K_a)h_u=g(K_a)g-\frac 1{2e}\{g(u)\otimes [g(u)](K)g-[g(u)](K)g\otimes
g(u)\}= \\ 
=N_{ij}.dx^i\wedge dx^j=N_{\alpha \beta }.e^\alpha \wedge e^\beta \text{ , }%
N_{ij}=-N_{ji}\text{ ,} \\ 
N_{ij}=\frac 12[g_{i\overline{k}}.g_{\overline{l}j}-g_{j\overline{k}}.g_{%
\overline{l}i}-\frac 1e(u_i.g_{\overline{l}j}-u_j.g_{\overline{l}i})u_{%
\overline{k}}]R^k\text{ }_{mnq}.u^mu^n.g^{ql}\text{ .}
\end{array}
\label{4.16}
\end{equation}

By means of the above expressions conditions can be found under which some
of the quantities $M$, $N$, $I$ vanish.

\subsubsection{Contravariant vector fields without rotation acceleration,
induced by the curvature ($N=0$)}

If the rotation acceleration $N$, induced by the curvature vanishes, i.e. if 
$N=0$, then the following proposition can be proved:

\textbf{Proposition 7.} The necessary and sufficient condition for the
existence of a contravariant vector field $u$ ($g(u,u)=e\neq 0$) without
rotation acceleration, induced by the curvature (i.e. with $N=0$) is the
condition 
\begin{equation}
K_a=\frac 1{2e}\{u\otimes [g(u)](K)-[g(u)](K)\otimes u\}\text{ .}
\label{4.17}
\end{equation}

Proof: 1. Sufficiency: From the above expression it follows 
\[
\begin{array}{c}
N=h_u(K_a)h_u= \\ 
=g(K_a)g-\frac 1{2e}\{g(u)\otimes [g(u)](K)g-[g(u)](K)g\otimes g(u)\}=0 
\text{ ,} \\ 
g([g(u)](K))=[g(u)](K)g\text{ .}
\end{array}
\]

2. Necessity: If $N=h_u(K_a)h_u=0$, then 
\[
\begin{array}{c}
g(K_a)g=\frac 1{2e}\{g(u)\otimes [g(u)](K)g-[g(u)](K)g\otimes g(u)\} \text{ ,%
} \\ 
K_a=\frac 1{2e}\{u\otimes [g(u)](K)-[g(u)](K)\otimes u\}\text{ .}
\end{array}
\]

In co-ordinate basis the necessary and sufficient condition has the forms 
\begin{equation}
\begin{array}{c}
K^{ij}=K^{ji}+\frac 1e.u_{\overline{l}}(u^i.K^{lj}-u^j.K^{li})\text{ ,} \\ 
\{R_{\overline{j}nim}-R_{\overline{i}mjn}-\frac 1e(u_{\overline{i}}R_{%
\overline{l}mnj}-u_{\overline{j}}R_{\overline{l}mni})u^l\}.u^mu^n=0\text{ , }
\end{array}
\label{4.18}
\end{equation}

\noindent where 
\[
R_{\overline{i}jkl}=g_{\overline{i}\overline{n}}.R^n\text{ }_{jkl}\text{ .} 
\]

\textbf{Proposition 8. }A sufficient condition for the existence of a
contravariant vector field $u$ ($g(u,u)=e\neq 0$) without rotation
acceleration, induced by the curvature (i.e. with $N=0$) is the condition 
\begin{equation}
K_a=0\text{ .}  \label{4.19}
\end{equation}

Proof: From $K_a=0$ and the form for $N$, $N=h_u(K_a)h_u$, it follows $N=0$.

In co-ordinate basis 
\begin{equation}
\begin{array}{c}
(R^i\text{ }_{klm}.g^{mj}-R^j\text{ }_{klm}.g^{mi})u^ku^l=0\text{ ,} \\ 
(R_{\overline{i}kjl}-R_{\overline{j}lik})u^ku^l=0\text{ .}
\end{array}
\label{4.20}
\end{equation}

$K_a=0$ can be presented also in the form 
\[
[g(\xi )]([R(u,v)]u)-[g(v)]([R(u,\xi )]u)=0\text{ , }\forall \xi ,v\in T(M) 
\text{ .} 
\]

In this case $M=G=g(K)g$ , $I=\overline{g}[G]$.

\textbf{Proposition 9.} A sufficient condition for the existence of a
contravariant vector field $u$ ($g(u,u)=e\neq 0$) without rotation
acceleration, induced by the curvature (i.e. with $N=0$) is the condition 
\begin{equation}
g(\eta ,[R(\xi ,v)]w)=g(\xi ,[R(\eta ,w)]v)\text{ , }\forall \eta ,\xi
,v,w\in T(M)\text{ ,}  \label{4.21}
\end{equation}

\noindent or in co-ordinate basis 
\begin{equation}
R_{\overline{i}jkl}=R_{\overline{k}lij}\text{ .}  \label{4.22}
\end{equation}

Proof: Because of $R(\xi ,u)=-R(u,\xi )$ and for $\eta =v$ the last
expression will be identical with the sufficient condition from proposition
9.

\subsubsection{Contravariant vector fields without shear acceleration $_sM$,
induced by the curvature ($_sM=0$)}

\textbf{Proposition 10.} The necessary and sufficient condition for the
existence of a contravariant vector field $u$ ($g(u,u)=e\neq 0$) without
shear acceleration, induced by the curvature (i.e. with $_sM=0$) is the
condition 
\begin{equation}
M=\frac 1{n-1}.I.h_u=\frac 1{n-1}.\overline{g}[M].h_u\text{ .}  \label{4.23}
\end{equation}

Proof: 1. Sufficiency: From the expression for $M$ and the definition of $%
_sM=M-\frac 1{n-1}.I.h_u$ it follows $_sM=0$.

2. Necessity: From $_sM=0=M-\frac 1{n-1}.I.h_u$ the form of $M$ follows.

In co-ordinate basis the necessary and sufficient condition can be written
in the form 
\begin{equation}
\begin{array}{c}
\{[g_{i\overline{k}}.g_{\overline{l}j}+g_{j\overline{k}}.g_{\overline{l}%
i}-\frac 1e(u_i.g_{\overline{l}j}+u_j.g_{\overline{l}i})u_{\overline{k}}]R^k%
\text{ }_{mns}g^{sl}- \\ 
-\frac 2{n-1}.R_{mn}(g_{ij}-\frac 1e.u_iu_j)\}u^mu^n=0\text{ .}
\end{array}
\label{4.24}
\end{equation}

The condition $_sM=0$ is identical with the condition for $K_s$: 
\begin{equation}
K_s=\frac 1{n-1}.I.h^u+\frac 1{2e}\{u\otimes [g(u)](K)+[g(u)](K)\otimes u\}%
\text{ .}  \label{4.25}
\end{equation}

\subsubsection{Contravariant vector fields without shear and expansion
acceleration, induced by the curvature ($_sM=0$, $I=0$)}

\textbf{Proposition 11.} A sufficient condition for the existence of a
contravariant vector field $u$ ($g(u,u)=e\neq 0$) without shear and
expansion acceleration, induced by the curvature (i.e. with $_sM=0$, $I=0$)
is the condition 
\begin{equation}
K_s=\frac 1{2e}\{u\otimes [g(u)](K)+[g(u)](K)\otimes u\}\text{ .}
\label{4.26}
\end{equation}

Proof: After acting on the left and on the right side of the last expression
with $g$%
\[
\begin{array}{c}
g(K_s)g=\frac 1{2e}\{g(u)\otimes [g(u)](K)g+[g(u)](K)g\otimes g(u)\} \text{ ,%
} \\ 
g([g(u)](K))=([g(u)](K))g=[g(u)](K)g\text{ , }u(g)=g(u)\text{ ,}
\end{array}
\]
and comparing the result with the form for $M$, 
\[
M=h_u(K_s)h_u=g(K_s)g-\frac 1{2e}\{g(u)\otimes [g(u)](K)g+[g(u)](K)g\otimes
g(u)\}\text{ ,} 
\]
it follows that $M=0$. Since $I=\overline{g}[M]$ it follows that $I=0$ and $%
_sM=0$.

\textbf{Proposition 12.} A sufficient condition for the existence of a
contravariant vector field $u$ ($g(u,u)=e\neq 0$) without shear and
expansion acceleration, induced by the curvature (i.e. with $_sM=0$, $I=0$)
is the condition 
\[
K_s=0\text{ .} 
\]

Proof: From the condition and the form of $M$, $M=h_u(K_s)h_u$, it follows
that $M=0$ and therefore $I=0$ and $_sM=0$.

\subsubsection{Contravariant vector fields without shear and rotation
acceleration, induced by the curvature ($_sM=0$, $N=0$)}

\textbf{Proposition 13.} A sufficient condition for the existence of a
contravariant vector field $u$ ($g(u,u)=e\neq 0$) without shear and rotation
acceleration, induced by the curvature (i.e. with $_sM=0$ , $N=0$) is the
condition 
\begin{equation}
\begin{array}{c}
\lbrack R(u,\xi )]v=\frac R{n(n-1)}[g(v,u).\xi -g(v,\xi ).u]\text{ , } \\ 
\forall v,\xi \in T(M)\text{ , }R\in C^r(M)\text{ .}
\end{array}
\label{4.27}
\end{equation}

Proof: Since $v$ is an arbitrary contravariant vector field it can be chosen
as $u$. Then, because of the relation 
\begin{equation}
h_u([R(u,\xi )]u)=h_u(K)h_u(\xi )=G(\xi )\text{ ,}  \label{4.28}
\end{equation}

\noindent it follows that 
\begin{equation}
G=h_u(K)h_u=\frac R{n(n-1)}.e.h_u=G_s\text{ , }G_a=h_u(K_a)h_u=0\text{ .}
\label{4.29}
\end{equation}

Therefore 
\begin{equation}
M=G_s=\frac{R.e}{n(n-1)}.h_u\text{ , }N=G_a=0\text{ , }I=\frac 1n.R.e\text{
, }_sM=0\text{ .}  \label{4.30}
\end{equation}

In co-ordinate basis the sufficient condition can be written in the form 
\begin{equation}
R^i\text{ }_{jkl}=\frac R{n(n-1)}(g_l^i.g_{\overline{j}\overline{k}%
}-g_k^i.g_{\overline{j}\overline{l}})  \label{4.31}
\end{equation}

\noindent and the following relations are fulfilled 
\begin{equation}
\begin{array}{c}
R_{jk}=R^l\text{ }_{jkl}=g_i^l.R^i\text{ }_{jkl}=\frac 1n.R.g_{\overline{j}%
\overline{k}}\text{ ,} \\ 
R=g^{jk}.R_{jk}\text{ ,} \\ 
I=R_{jk}.u^ju^k=\frac 1n.R.e\text{ .}
\end{array}
\label{4.32}
\end{equation}

\textbf{Proposition 14.} The necessary and sufficient conditions for the
existence of $K$ in the form 
\begin{equation}
K=\frac 1{n-1}.g[K].h^u  \label{4.33}
\end{equation}

\noindent are the conditions 
\[
_sM=0\text{ , }K_a=0\text{ .} 
\]

Proof: 1. Sufficiency: From $K_a=0$ it follows that $K=K_s$, $N=0$ and $%
M=g(K_s)g=g(K)g$. Therefore, $I=\overline{g}[M]=g[K]$. From $_sM=M-\frac
1{n-1}.I.h_u=0$ it follows that $M=\frac 1{n-1}.g[K].h_u=g(K)g$. From the
last expression it follows the above condition for $K$.

2. Necessity: From the condition $K=\frac 1{n-1}.g[K].h^u$ it follows that $%
K=K_s$ and therefore $K_a=0$, $N=0$ and $M=\frac 1{n-1}.g[K].h_u$, $I=g[K]$
(because of $h_u(h^u)h_u=h_u$, $h_u(\overline{g})h_u=h_u$). From the forms
of $M$ and $I$ it follows that $_sM=0$.

\textbf{Proposition 15.} A sufficient condition for the existence of a
contravariant vector field $u$ ($g(u,u)=e\neq 0$) without shear and rotation
acceleration, induced by the curvature (i.e. with $_sM=0$ , $N=0$) is the
condition 
\[
K=\frac 1{n-1}.g[K].h^u\text{ .} 
\]

Proof: Follows immediately from proposition 15.

\subsubsection{Contravariant vector fields without expansion acceleration,
induced by the curvature ($I=0$)}

By means of the covariant metric $g$ and the tensor field $K(v,\xi )$ the
notion contravariant Ricci tensor $Ricci$ can be introduced 
\begin{equation}
Ricci(v,\xi )=g[K(v,\xi )]\text{ , }\forall v,\xi \in T(M)\text{ ,}
\label{4.34}
\end{equation}

\noindent where 
\begin{equation}
K(v,\xi )=R^i\text{ }_{jkl}.g^{lm}.v^j.\xi ^k.\partial _i\otimes \partial
_m=R^\alpha \text{ }_{\beta \gamma \kappa }.g^{\kappa \delta }.v^\beta .\xi
^\gamma .e_\alpha \otimes e_\delta \text{ ,}  \label{4.35}
\end{equation}

\noindent and the following relations are fulfilled 
\begin{equation}
\begin{array}{c}
Ricci(e_\alpha ,e_\beta )=g[K(e_\alpha ,e_\beta )]=R_{\alpha \beta }\text{ ,}
\\ 
Ricci(\partial _i,\partial _j)=g[K(\partial _i,\partial _j)]=R_{ij}\text{ ,}
\\ 
Ricci(u,u)=g[K(u,u)]=g[K]=I\text{ .}
\end{array}
\label{4.36}
\end{equation}

\textbf{Proposition 16.} The necessary and sufficient condition for the
existence of a contravariant vector field $u$ ($g(u,u)=e\neq 0$) without
expansion acceleration, induced by the curvature (i.e. with $I=0$) is the
condition 
\[
Ricci(u,u)=0\text{ .} 
\]

Proof: It follows immediately from the relation $Ricci(u,u)=g[K(u,u)]=g[K]=I$%
.

\textbf{Proposition 17.} A sufficient condition for the existence of a
contravariant vector field $u$ ($g(u,u)=e\neq 0$) without expansion
acceleration, induced by the curvature (i.e. with $I=0$) is the condition 
\begin{equation}
\begin{array}{c}
Ricci(e_\alpha ,e_\beta )=R_{\alpha \beta }=R^\gamma \text{ }_{\alpha \beta
\gamma }=0\text{ ,} \\ 
Ricci(\partial _i,\partial _j)=R_{ij}=R^l\text{ }_{ijl}=0\text{ .}
\end{array}
\label{4.37}
\end{equation}

Proof: From $Ricci(\partial _i,\partial _j)=R_{ij}=0$ it follows that 
\[
R_{ij}.u^iu^j=u^iu^j.Ricci(\partial _i,\partial _j)=Ricci(u,u)=I=0. 
\]

In non-co-ordinate basis the proof is analogous to that in co-ordinate basis.

The existence of contravariant vector fields with vanishing characteristics,
induced by the curvature, is important for mathematical models of
gravitational interactions in theories over ($\overline{L}_n,g$)-spaces.

\section{Conclusion}

The covariant and contravariant metric introduced over differentiable
manifolds with contravariant and covariant affine connection allow
applications for mathematical models of dynamic systems described over $( 
\overline{L}_n,g)$-spaces. On the other side different type of geometries
can be considered by imposing certain additional conditions of the type of
metric transport on the metric. Additional conditions determined by
different ''draggings along'' of the metric can have physical interpretation
connected with changes of the length of a vector field and with changes of
the angle between two vector fields.

The introduction of contravariant and covariant projective metric
corresponding to a non-isotropic (non-null) contravariant vector field
allows the evolution of tensor analysis over sub-manifolds of a manifold
with contravariant and covariant connection and metric and its applications
for descriptions of the evolution of physical systems over $(\overline{L}%
_n,g)$-spaces.

The kinematic characteristics, connected with the introduced notions
relative velocity and relative acceleration can be used for description of
different dynamic systems by means of mathematical models, using
differentiable manifold $M$ with contravariant and covariant affine
connection and metric as a model of space-time ($\dim M=4$)(ETG in $V_n$%
-spaces, Einstein-Cartan theory in $U_n$-spaces), or as a model for the
consideration of dynamic characteristics of some physical systems (theories
of the type of Kaluza-Klein in $V_n$-spaces ($n\succ 4$), relativistic
hydrodynamics etc.). At the same time the kinematic characteristics can be
used for a more correct formulation of problems, connected with the
experimental check-up of modern gravitational theories.

In the case of general relativity theory proposition 5. can be used for
describing the characteristics of gravitational detectors: If test particles
are considered to move in an external gravitational field ($R_{ij}=0$), then
their relative acceleration will be caused only by the curvature shear
acceleration. Therefore, gravitational wave detectors have to be able to
detect accelerations of the type of shear acceleration (and not of the type
of expansion acceleration), if the energy-momentum tensor of the detector is
neglected as a source of a gravitational field.

\textbf{References}

\begin{enumerate}
\item  Barvinskii A. O., Pnomariev V. N., Obukhov J.N. \textbf{1985 } 
\textit{Geometric-dynamic methods and gauge approach in the theory of
gravitational interactions.} Energoatomizdat, Moscow (in russ.)

\item  Bishop R. L., Goldberg S.I. \textbf{1968 }\textit{Tensor Analysis on
Manifolds.} The Macmillan Company, New York

\item  Boothby W.M. \textbf{1975} \textit{An Introduction to Differentiable
Manifolds and Riemannian Geometry}. Academic Press,New York

\item  Choquet-Bruhat Y., DeWitt-Morette C., Dillard-Bleik M. \textbf{1977 } 
\textit{Analysis, Manifolds and Physics}. North-Holland Pub. Co., Amsterdam

\item  Eddington A.S. \textbf{1925 } \textit{Relativitaetstheorie in
mathematischer Behandlung}. Verlag von Julius Springer, Berlin

\item  Efimov N. B., Rosendorn E. R. \textbf{1974 }\textit{Linear Algebra
and Multidimensional Geometry.} 2-nd Ed. Nauka, Moscow (in russ.)

\item  Ehlers J. \textbf{1961 }Abhandlungen d. Mainzer Akademie d.
Wissenschaften, Math.-Naturwiss. Kl. Nr.$\underline{11}$ ,

\item  Greenberg Ph.J. \textbf{1970} J. Math. Ann. Apply., $\underline{30}$
, 128

\item  Greub W. \textbf{1978 }\textit{Multilinear Algebra}. Springer Verlag,
New York

\item  Greub W., Halperin St., Vanstone R. \textbf{1972 }\textit{\
Connections, Curvature, and Cohomology}. Vol.I., Academic Press, New York
and London

\item  .................\textbf{1973} \textit{Connections, Curvature, and
Cohomology}. Vol.II., Academic Press, New York and London

\item  Hawking S.W., Ellis G. \textbf{1973 } \textit{The Large Scale
Structure of Space-Time}.Cambridge U.P., Cambridge , \#\# 4.1.,4.4

\item  Hecht R.D., Hehl F.W. \textbf{1991 }Proc. 9th Italian Conf. on Gen.
Rel. and Grav. Physics. Capri, Italy 1991. World Sci. Pub. Co., Singapore
,pp.246-291

\item  Hehl F.W. \textbf{1966 } Dissertation. TU Clausthal 1966. Abh. d.
Braunschweigischen Wiss. Gesellschaft\textbf{\ 18}, 98- 130

\item  .................\textbf{1970} Habilitationsschrift. TU Clausthal .
S.1-78

\item  .................\textbf{1973} Gen. Rel. and Grav. \textbf{4}, 4,
333-349

\item  .................\textbf{1974} Preprint. Princeton Univ.

\item  Hehl F.W., Kerlick G.D. \textbf{1978} Gen. Rel. and Grav.\textbf{\ 9}%
, 8, 691

\item  Hehl F.W., von der Heyde P. \textbf{1973 } Ann. Inst. Henri
Poincare'. Sec. A. \textbf{19}, 2, 179-196

\item  Iliev B.Z. \textbf{1992 } Comm. JINR E5-92-507, Dubna, pp. 1-17

\item  ...............\textbf{1992 }Comm. JINR E5-92-508, Dubna, pp. 1-16

\item  ...............\textbf{1992}Comm. JINR E5-92-543, Dubna, pp. 1-15

\item  Iliev B., Manoff S. \textbf{1983} Comm. JINR P2-83-897, Dubna, pp.
1-16

\item  Ivanenko D.D., Pronin P.I., Sardanashvily G.A. \textbf{1985} \textit{%
Gauge theory of gravitation}. Izd. Moskovskogo universiteta, Moscow (in
russ.)

\item  Kobayashi S., Nomizu K. \textbf{1963 }\textit{Foundations of
Differential Geometry.} Vol. I. Interscience Publishers., New York

\item  Kramer D., Stephani H. \textbf{1983 } In Proc. 9th Intern. Conf. in
General Relativity and Gravitation. VEB Deutscher Verlag der Wissenschaften,
Berlin ,S.79-80

\item  Kramer D., Stephani H., MacCallum M., Herlt E. \textbf{1980 } \textit{%
Exact Solutions of Einstein's Field Equations}.VEB Deutscher Verlag der
Wissenschaften, Berlin

\item  Lichnerowicz A. \textbf{1979 } In \textit{Astrofisica e cosmologia
gravitazione, quanti e relativita}. Giunti Barbera, Firenze

\item  Lightman A.P., Press W.H., Price R.H., Teukolsky S.A.\textbf{\ 1975 } 
\textit{Problem Book in Relativity and Gravitation}. Princeton Univ. Press,
Princeton, New Jersey

\item  Logunov A.A., Mestvirishvili M.A. \textbf{1989 } \textit{Relativistic
theory of gravitation.} Nauka, Moscow (in russ.)

\item  Lovelock D., Rund H. \textbf{1975 } Tensors, Differential Forms, and
Variational Principles. John Wiley \& Sons, New York

\item  Manoff S..\textbf{1979} Gen. Rel. and Grav. \textbf{11,} 189-204

\item  ................\textbf{1984 }6th Sov. Grav. Conf. Contr. Papers.
Moscow, pp.231-232

\item  ................\textbf{1985 }In \textit{Gravitational Waves. }%
JINR-Dubna, P2-85-667, Dubna, pp.157-168

\item  ................\textbf{1986 }11th Intern. Conf. on Gen. Rel. and
Grav. Contr. Papers. Stockholm

\item  ................\textbf{1987 }Comm. JINR E2-87-679, Dubna , c.1-16

\item  ................\textbf{1989} In Proc. of the Workshop on Development
and Building a Generator and Detector for Gravitational Waves. JINR Dubna
D4-89-221, Dubna, c. 42-54

\item  ................\textbf{1989} 12th Intern. Conf. on Gen. Rel. and
Grav.,Colorado of Boulder, USA,. Univ. Colorado, Colorado 1989, vol.1.,p.95

\item  ................\textbf{1989} 12th Intern. Conf. on Gen. Rel. and
Grav.,Colorado of Boulder, USA, 1989. Univ. Colorado, Colorado 1989, vol.
1.,p.178

\item  ................\textbf{1991} J. Math. Phys. \textbf{32}, 3, 728-734

\item  ................\textbf{1991} Comm. JINR Dubna, E2-91-77, Dubna,
pp.1-16

\item  ................\textbf{1991} Comm. JINR Dubna, E2-91-78, Dubna,
pp.1-15

\item  ................\textbf{1991} Proc. 9th Italian Conf. on Gen. Rel.
and Grav. Physics. Capri, Italy 1991. World Sci. Publ. Co., Singapore,
pp.476-480

\item  ................\textbf{1992} Comm. JINR Dubna, E2-92-19, Dubna,
pp.1-12

\item  ................\textbf{1992 }13th Intern. Conf. on Gen. Rel. and
Grav., Huerta Grande, Cordoba, Argentina 1992. Contr. Papers

\item  Matsushima Y. \textbf{1972 } \textit{Differentiable Manifolds}.
Marcel Dekker, Inc.,New York

\item  Norden A.P. \textbf{1976 } \textit{Spaces with affine connection}.
2-nd Ed., Nauka, Moscow (in russ.)

\item  Rosen N. \textbf{1973 } Gen. Rel. and Grav.,\textbf{\ 4}, 6, 435-447

\item  ...............\textbf{1974 }Ann. of Phys., \textbf{84,}1-2, 455-473

\item  Schmutzer E. \textbf{1968 } \textit{Relativistische Physik
(Klassische Theorie)}. B.G. Teubner Verlagsgesellschaft, Leipzig

\item  Schouten J. A. \textbf{1951 }\textit{Tensor Analysis for Physicist.}
Clarendon Press, Oxford

\item  Schroedinger E. \textbf{1950 }\textit{Space-Time Structure. }%
Cambridge at the University Press, Cambridge

\item  Slebodzinski W. \textbf{1931 } Bull. Acad. Roy. Belgique. \textbf{17}%
, 864

\item  Stephani H. \textbf{1977} \textit{Allgemeine Relativitaetstheorie.}
VEB Deutscher Verlag d. Wiss., Berlin, pp.163-166

\item  Swaminarayan N.S., Safko J.L. \textbf{1983 } J. Math. Phys. \textbf{%
24, }4, 883-885

\item  Synge J.L. \textbf{1960} \textit{Relativity: the general theory}.
North-Holland Publ. Co., Amsterdam, Ch IV., \#3.

\item  Tchernikov N.A. \textbf{1987 } In Elementary particles and atomic
nucleus. \textbf{18,} 5, 1000-1034 (in russ.)

\item  ...........................\textbf{1988} Preprint JINR P2-88-778,
Dubna (in russ.)

\item  ...........................\textbf{1990} Comm. JINR P2-90-399, Dubna,
pp. 1-17 (in russ.)

\item  von der Heyde P.\textbf{\ 1975 } Lett. Nuovo Cim. \textbf{14,} 7,
250-252

\item  Yano K. \textbf{1957 } \textit{The Theory of Lie Derivatives and its
Applications}. North-Holland Pub. Co., Amsterdam
\end{enumerate}

\appendix 

\section{Kinematic characteristics connected with the relative acceleration
and expressed in terms of the kinematic characteristics connected with the
relative velocity}

The deformation, shear, rotation and expansion acceleration can be expressed
in terms of the shear, rotation and expansion velocity.

a) Deformation acceleration tensor $A$: 
\begin{equation}  \label{A.1}
\begin{array}{c}
A=\frac 1eh_u(a)\otimes h_u(a)+\sigma ( \overline{g})\sigma +\omega (%
\overline{g})\omega +\frac 2{n-1}.\theta .(\sigma +\omega )+\frac
1{n-1}(\theta ^{.}+\frac{\theta ^2}{n-1})h_u+ \\ 
+\sigma ( \overline{g})\omega +\omega (\overline{g})\sigma +\nabla _u\sigma
+\nabla _u\omega +\frac 1e.h_u(a)\otimes (g(u))(2k-\nabla _u\overline{g})h_u+
\\ 
+\frac 1e[\sigma (a)\otimes g(u)+g(u)\otimes \sigma (a)]+\frac 1e[\omega
(a)\otimes g(u)-g(u)\otimes \omega (a)]+ \\ 
+h_u(\nabla _u\overline{g})\sigma +h_u(\nabla _u\overline{g})\omega \text{ ,}
\end{array}
\end{equation}

\noindent where 
\[
k=\epsilon +s-(m+q)=k_0-(m+q)\text{ ,} 
\]
\begin{equation}
k(g)\pounds _\xi u=\nabla _{\pounds _\xi u}u-T(\pounds _\xi u,u)\text{ .}
\label{A.2}
\end{equation}

In index form 
\begin{equation}  \label{A.3}
\begin{array}{c}
A_{ij}=\frac 1e.h_{i \overline{k}}a^ka^lh_{\overline{l}j}+\sigma _{ik}g^{%
\overline{k}\overline{l}}\sigma _{lj}+\omega _{ik}g^{\overline{k}\overline{l}%
}\omega _{lj}+\frac 2{n-1}.\theta .\sigma _{ij}+\frac 1{n-1}(\theta ^{.}+%
\frac{\theta ^2}{n-1})h_{ij}+ \\ 
+\sigma _{ij;k}u^k+\frac 1e.a^k[\sigma _{i \overline{k}}u_j+\sigma _{j%
\overline{k}}u_i+h_{\overline{k}(i}h_{j)\overline{l}}u_{\overline{n}%
}(2k^{nl}-g^{nl}\text{ }_{;r}u^r)]+ \\ 
+\frac 12(h_{i \overline{k}}g^{kl}\text{ }_{;r}u^r\sigma _{\overline{l}%
j}+h_{j\overline{k}}g^{kl}\text{ }_{;r}u^r\sigma _{\overline{l}i})+\frac
12(h_{i\overline{k}}g^{kl}\text{ }_{;r}u^r\omega _{\overline{l}j}+h_{j%
\overline{k}}g^{kl}\text{ }_{;r}u^r\omega _{\overline{l}i})+ \\ 
+\sigma _{ik}g^{\overline{k}\overline{l}}\omega _{lj}-\sigma _{jk}g^{%
\overline{k}\overline{l}}\omega _{li}+\frac 2{n-1}.\theta .\omega
_{ij}+\omega _{ij;r}u^r+ \\ 
+\frac 1e.a^k[\omega _{i \overline{k}}u_j-\omega _{j\overline{k}}u_i+h_{%
\overline{k}[i}h_{j]\overline{l}}u_{\overline{n}}(2k^{nl}-g^{nl}\text{ }%
_{;r}u^r)]+ \\ 
+\frac 12(h_{i \overline{k}}g^{kl}\text{ }_{;r}u^r\sigma _{\overline{l}%
j}-h_{j\overline{k}}g^{kl}\text{ }_{;r}u^r\sigma _{\overline{l}i})+\frac
12(h_{i\overline{k}}g^{kl}\text{ }_{;r}u^r\omega _{\overline{l}j}-h_{j%
\overline{k}}g^{kl}\text{ }_{;r}u^r\omega _{\overline{l}i})= \\ 
=D_{ij}+W_{ij}\text{ ,}
\end{array}
\end{equation}
\[
A_{(ij)}=\frac 12(A_{ij}+A_{ji})\text{ , }A_{[ij]}=\frac 12(A_{ij}-A_{ji)} 
\text{ .} 
\]

b) Shear acceleration tensor $_sD=D-\frac 1{n-1}.U.h_u$: 
\begin{equation}  \label{A.4}
\begin{array}{c}
D=\frac 1eh_u(a)\otimes h_u(a)+\sigma ( \overline{g})\sigma +\omega (%
\overline{g})\omega + \\ 
+\frac 2{n-1}.\theta .\sigma +\frac 1{n-1}(\theta ^{.}+ \frac{\theta ^2}{n-1}%
)h_u+\nabla _u\sigma + \\ 
+\frac 1{2e}[h_u(a)\otimes (g(u))(2k-\nabla _u \overline{g}%
)h_u+h_u((g(u))(2k-\nabla _u\overline{g}))\otimes h_u(a)]+ \\ 
+\frac 1e[\sigma (a)\otimes g(u)+g(u)\otimes \sigma (a))]+ \\ 
+\frac 12[h_u(\nabla _u\overline{g})\sigma +\sigma (\nabla _u\overline{g}%
)h_u]+\frac 12[h_u(\nabla _u\overline{g})\omega -\omega (\nabla _u\overline{g%
})h_u]\text{ .}
\end{array}
\end{equation}

In index form 
\begin{equation}  \label{A.5}
\begin{array}{c}
D_{ij}=D_{ji}=\frac 1e.h_{i \overline{k}}a^ka^lh_{\overline{l}j}+\sigma _{i%
\overline{k}}g^{kl}\sigma _{\overline{l}j}+\omega _{i\overline{k}%
}g^{kl}\omega _{\overline{l}j}+ \\ 
+\frac 2{n-1}.\theta .\sigma _{ij}+\frac 1{n-1}(\theta ^{.}+ \frac{\theta ^2%
}{n-1})h_{ij}+\sigma _{ij;k}u^k+ \\ 
+\frac 1e.a^k\{\sigma _{i \overline{k}}u_j+\sigma _{j\overline{k}}u_i+ \\ 
+h_{\overline{k}(i}h_{j)\overline{l}}[g^{ml}(e_{,m}-g_{rs;m}u^{\overline{r}%
}u^{\overline{s}}-2T_{mr}^nu^ru_{\overline{n}})-u_{\overline{n}}g^{nl}\text{ 
}_{;m}u^m]\}+ \\ 
+\frac 12(h_{ik}g^{\overline{k}\overline{l}}\text{ }_{;m}u^m\sigma
_{lj}+h_{jk}g^{\overline{k} \overline{l}}\text{ }_{;m}u^m\sigma _{li})+ \\ 
+\frac 12(h_{ik}g^{\overline{k}\overline{l}}\text{ }_{;m}u^m\omega
_{lj}+h_{jk}g^{\overline{k}\overline{l}} \text{ }_{;m}u^m\omega _{li})\text{
.}
\end{array}
\end{equation}

c) Rotation acceleration tensor $W$%
\begin{equation}  \label{A.6}
\begin{array}{c}
=\sigma ( \overline{g})\omega +\omega (\overline{g})\sigma +\frac
2{n-1}.\theta .\omega +\nabla _u\omega + \\ 
+\frac 1e[\omega (a)\otimes g(u)-g(u)\otimes \omega (a)]+ \\ 
+\frac 1{2e}[h_u(a)\otimes (g(u))(2k-\nabla _u \overline{g}%
)h_u-h_u((g(u))(2k-\nabla _u\overline{g}))\otimes h_u(a)]+ \\ 
+\frac 12[h_u(\nabla _u\overline{g})\sigma -\sigma (\nabla _u\overline{g}%
)h_u]+\frac 12[h_u(\nabla _u\overline{g})\omega +\omega (\nabla _u\overline{g%
})h_u]\text{ .}
\end{array}
\end{equation}

In index form 
\begin{equation}  \label{A.7}
\begin{array}{c}
W_{ij}=-W_{ji}=\sigma _{ik}g^{\overline{k}\overline{l}}\omega _{lj}-\sigma
_{jk}g^{\overline{k}\overline{l}}\omega _{li}+\frac 2{n-1}.\theta .\omega
_{ij}+\omega _{ij;k}u^k+ \\ 
+\frac 1ea^k\{\omega _{i \overline{k}}u_j-\omega _{j\overline{k}}u_i+h_{%
\overline{k}[i}h_{j]\overline{l}}[g^{ml}(e_{,m}-g_{rs;m}u^{\overline{r}}u^{%
\overline{s}}- \\ 
-2T_{mr}^nu^ru_{\overline{n}})-u_{\overline{n}}g^{nl}\text{ }_{;m}u^m]\}+ \\ 
+\frac 12(h_{ik}g^{\overline{k}\overline{l}}\text{ }_{;m}u^m\sigma
_{lj}-h_{jk}g^{\overline{k} \overline{l}}\text{ }_{;m}u^m\sigma _{li})+ \\ 
+\frac 12(h_{i\overline{k}}g^{kl}\text{ }_{;m}u^m\omega _{\overline{l}j}-h_{j%
\overline{k}}g^{kl}\text{ }_{;m}u^m\omega _{\overline{l}i})\text{ .}
\end{array}
\end{equation}

d) Expansion acceleration $U$%
\begin{equation}  \label{A.8}
\begin{array}{c}
U=\frac 1e.g(a,a)+ \overline{g}[\sigma (\overline{g})\sigma ]+\overline{g}[%
\omega (\overline{g})\omega ]+\theta ^{.}+\frac 1{n-1}.\theta ^2+ \\ 
+\frac 1e[2g(u,\nabla _au)-2g(u,T(a,u))+(\nabla _ug)(a,u)] \\ 
-\frac 1{e^2}.g(u,a).[3g(u,a)+(\nabla _ug)(u,u)]\text{ .}
\end{array}
\end{equation}

In index form 
\begin{equation}  \label{A.9}
\begin{array}{c}
U=\frac 1e.g_{\overline{i}\overline{j}}a^ia^j+g^{\overline{i}\overline{j}}g^{%
\overline{k} \overline{l}}\sigma _{ik}\sigma _{jl}-g^{\overline{i}\overline{j%
}}g^{\overline{k}\overline{l}}\omega _{ik}\omega _{jl}+\theta ^{.}+\frac
1{n-1}.\theta ^2+ \\ 
+\frac 1e.g_{\overline{k}\overline{l}}a^k[g^{ml}(e_{,m}-g_{rs;m}u^{\overline{%
r}}u^{\overline{s}}-2T_{mr}^nu^ru_{\overline{n}})-u_{\overline{n}}g^{nl}%
\text{ }_{;m}u^m]- \\ 
+\frac 1{e^2}[\frac 34(e_{,k}u^k)^2-(e_{,l}u^l)g_{ij;k}u^ku^{\overline{i}}u^{%
\overline{j}}+\frac 14(g_{ij;k}u^ku^{\overline{i}}u^{\overline{j}})^2]\text{
.}
\end{array}
\end{equation}

e) Torsion-free and curvature-free shear acceleration tensor $_{sF}D_0$ 
\[
_{sF}D_0=_FD_0-\frac 1{n-1}._FU_0.h_u 
\]
\begin{equation}  \label{A.10}
_FD_0=h_u(b_s)h_u\text{ , }_FU_0=g[b]-\frac 1e.g(u,\nabla _ua) \text{ .}
\end{equation}

In index form 
\begin{equation}  \label{A.11}
(_FD_0)_{ij}=(_FD_0)_{ji}=\frac 12.h_{i\overline{k}}(a^k\text{ }%
_{;n}g^{nl}+a^l\text{ }_{;n}g^{nk})h_{\overline{l}j}\text{ ,}
\end{equation}
\begin{equation}  \label{A.12}
\begin{array}{c}
_FU_0=a^k \text{ }_{;k}-\frac 1e.g_{\overline{k}\overline{l}}u^ka^l\text{ }%
_{;m}u^m= \\ 
=a^k\text{ }_{;k}-\frac 1e[(g_{\overline{k}\overline{l}%
}u^ka^l)_{;m}u^m-g_{kl;m}u^mu^{\overline{k}}a^{\overline{l}}-g_{\overline{k} 
\overline{l}}a^ka^l]\text{ .}
\end{array}
\end{equation}

f) Torsion-free and curvature-free rotation acceleration tensor $_FW_0$%
\begin{equation}  \label{A.13}
_FW_0=h_u(b_a)h_u\text{ .}
\end{equation}

In index form 
\begin{equation}  \label{A.14}
(_FW_0)_{ij}=-(_FW_0)_{ji}=\frac 12.h_{i\overline{k}%
}(a_{;n}^kg^{nl}-a_{;n}^lg^{nk})h_{\overline{l}j}.
\end{equation}

g) Torsion-free and curvature-free expansion acceleration $_FU_0$ (s. e)).

h) Curvature-free shear acceleration tensor $_sD_0=D_0-\frac 1{n-1}.U_0.h_u$%
\begin{equation}  \label{A.15}
\begin{array}{c}
D_0=h_u(b_s)h_u-\frac 12[_sP( \overline{g})\sigma +\sigma (\overline{g}%
)_sP]-\frac 12[Q(\overline{g})\omega +\omega (\overline{g})Q]- \\ 
-\frac 1{n-1}(\theta _1.\sigma +\theta ._sP)-\frac 1{n-1}(\theta ^{.}+\frac
1{n-1}.\theta _1.\theta )h_u-\nabla _u(_sP)- \\ 
-\frac 12[_sP( \overline{g})\omega -\omega (\overline{g})_sP]-\frac 12[Q(%
\overline{g})\sigma -\sigma (\overline{g})Q]- \\ 
-\frac 1{2e}[h_u(a)\otimes (g(u))(m+q)h_u+h_u((g(u))(m+q))\otimes h_u(a)- \\ 
-\frac 1e[_sP(a)\otimes g(u)+g(u)\otimes _sP(a)]- \\ 
-\frac 12[h_u(\nabla _u\overline{g})_sP+_sP(\nabla _u\overline{g})h_u]-\frac
12[h_u(\nabla _u\overline{g})Q-Q(\nabla _u\overline{g})h_u]\text{ .}
\end{array}
\end{equation}

In index form 
\begin{equation}
\begin{array}{c}
(D_0)_{ij}=(D_0)_{ji}=h_{\overline{k}(i}h_{j)\overline{l}}a^k\text{ }%
_{;m}g^{ml}-_sP_{k(i}\sigma _{j)l}g^{\overline{k}\overline{l}}-Q_{k(i}\omega
_{j)l}g^{\overline{k}\overline{l}}- \\ 
-\frac 1{n-1}(\theta _1.\sigma _{ij}+\theta ._sP_{ij})-\frac 1{n-1}(\theta
_1^{.}+\frac 1{n-1}.\theta _1.\theta )h_{ij}- \\ 
-_sP_{ij;m}u^m+_sP_{k(i}\omega _{j)l}g^{\overline{k}\overline{l}%
}+Q_{k(i}\sigma _{j)l}g^{\overline{k}\overline{l}}- \\ 
-\frac 1e.a^k[_sP_{i\overline{k}}u_j+_sP_{j\overline{k}}u_i+h_{\overline{k}%
(i}h_{j)\overline{l}}u_{\overline{n}}T_{mr}^nu^rg^{ml}]- \\ 
-_sP_{\overline{k}(i}h_{j)\overline{l}}g^{kl}\text{ }_{;m}u^m-Q_{\overline{k}%
(i}h_{j)\overline{l}}g^{kl}\text{ }_{;m}u^m\text{ .}
\end{array}
\label{A.16}
\end{equation}

i) Curvature-free rotation acceleration tensor $W_0$%
\begin{equation}  \label{A.17}
\begin{array}{c}
W_0=h_u(b_a)h_u-\frac 12[_sP( \overline{g})\sigma -\sigma (\overline{g}%
)_sP]-\frac 12[Q(\overline{g})\omega -\omega (\overline{g})Q]- \\ 
-\frac 1{n-1}(\theta _1.\omega +\theta .Q)-\nabla _uQ-\frac 12[_sP( 
\overline{g})\omega +\omega (\overline{g})_sP]- \\ 
-\frac 12[Q( \overline{g})\sigma +\sigma (\overline{g})Q]-\frac
1e[Q(a)\otimes g(u)-g(u)\otimes Q(a)]- \\ 
-\frac 1{2e}[h_u(a)\otimes (g(u))(m+q)h_u-h_u((g(u))(m+q))\otimes h_u(a)]-
\\ 
-\frac 12[h_u(\nabla _u\overline{g})_sP-_sP(\nabla _u\overline{g})h_u]-\frac
12[h_u(\nabla _u\overline{g})Q+Q(\nabla _u\overline{g})h_u]\text{ .}
\end{array}
\end{equation}

In index form 
\begin{equation}  \label{A.18}
\begin{array}{c}
(W_0)_{ij}=-(W_0)_{ji}=h_{\overline{k}[i}h_{j]\overline{l}}a^k\text{ }%
_{;m}g^{ml}-_sP_{\overline{k}[i}\sigma _{j]\overline{l}}g^{kl}-Q_{\overline{k%
}[i}\omega _{j]\overline{l}}g^{kl}- \\ 
-\frac 1{n-1}(\theta _1.\omega _{ij}+\theta .Q_{ij})-Q_{ij;m}u^m+ \\ 
+_sP_{k[i}\omega _{j]l}g^{\overline{k}\overline{l}}+Q_{k[i}\sigma _{j]l}g^{%
\overline{k}\overline{l}}- \\ 
-\frac 1e.a^k(Q_{i \overline{k}}u_j-Q_{j\overline{k}}u_i+h_{\overline{k}[%
i}h_{j]\overline{l}}u_{\overline{n}}T_{mr}^nu^rg^{ml})+ \\ 
+_sP_{\overline{k}[i}h_{j]\overline{l}}g^{kl}\text{ }_{;m}u^m+Q_{\overline{k%
}[i}h_{j]\overline{l}}g^{kl}\text{ }_{;m}u^m\text{ .}
\end{array}
\end{equation}

j) Curvature-free expansion acceleration $U_0$%
\begin{equation}  \label{A.19}
\begin{array}{c}
U_0=g[b]- \overline{g}[_sP(\overline{g})\sigma ]-\overline{g}[Q(\overline{g}%
)\omega ]-\theta _1^{.}-\frac 1{n-1}.\theta _1.\theta - \\ 
-\frac 1e[g(u,T(a,u))+g(u,\nabla _ua)]\text{ .}
\end{array}
\end{equation}

In index form 
\begin{equation}  \label{A.20}
\begin{array}{c}
U_0=a^k \text{ }_{;k}-g^{\overline{i}\overline{j}}\text{ }._sP_{ik}g^{%
\overline{k} \overline{l}}\sigma _{lj}-g^{\overline{i}\overline{j}}Q_{ik}g^{%
\overline{k} \overline{l}}\omega _{lj}-\theta _1^{.}-\frac 1{n-1}.\theta
_1.\theta - \\ 
-\frac 1e[a^k(u_{\overline{n}}T_{km}^nu^m-2g_{\overline{k}\overline{m}%
;l}u^lu^m-g_{\overline{k}\overline{l}}a^l)+ \\ 
+\frac 12(e_{,k}u^k)_{,l}u^l-\frac 12(g_{mn;r}u^r)_{;s}u^su^{\overline{m}}u^{%
\overline{n}}]\text{ .}
\end{array}
\end{equation}

k) Shear acceleration tensor, induced by the torsion, $_{sT}D_0$%
\[
_{sT}D_0=_TD_0-\frac 1{n-1}._TU_0.h_u 
\]

\begin{equation}  \label{A.21}
\begin{array}{c}
_TD_0=\frac 12[_sP( \overline{g})\sigma +\sigma (\overline{g})_sP]+\frac
12[Q(\overline{g})\omega +\omega (\overline{g})Q]+ \\ 
+\frac 1{n-1}(\theta _1.\sigma +\theta ._sP)+\frac 1{n-1}(\theta
_1^{.}+\frac 1{n-1}.\theta _1.\theta )h_u+\nabla _u(_sP)+ \\ 
+\frac 12[_sP( \overline{g})\omega -\omega (\overline{g})_sP]+\frac 12[Q(%
\overline{g})\sigma -\sigma (\overline{g})Q]+ \\ 
+\frac 1{2e}[h_u(a)\otimes (g(u))(m+q)h_u+h_u((g(u))(m+q))\otimes h_u(a)]+
\\ 
+\frac 1e[_sP(a)\otimes g(u)+g(u)\otimes _sP(a)]+ \\ 
+\frac 12[h_u(\nabla _u\overline{g})_sP+_sP(\nabla _u\overline{g})h_u]+\frac
12[h_u(\nabla _ug)Q-Q(\nabla _ug)h_u]\text{ .}
\end{array}
\end{equation}

In index form 
\begin{equation}  \label{A.22}
(_TD_0)_{ij}=(_FD_0)_{ij}-(D_0)_{ij}\text{ .}
\end{equation}

l) Expansion acceleration, induced by the torsion, $_TU_0$%
\begin{equation}  \label{A.23}
_TU_0=\overline{g}[_sP(\overline{g})\sigma ]+\overline{g}[Q( \overline{g}%
)\omega ]+\theta _1^{.}+\frac 1{n-1}.\theta _1.\theta +\frac 1e.g(u,T(a,u))%
\text{ .}
\end{equation}

In index form 
\[
_TU_0=_FU_0-U_0\text{ .} 
\]

m) Rotation acceleration tensor, induced by the torsion, $_TW_0$%
\begin{equation}  \label{A.24}
\begin{array}{c}
_TW_0=\frac 12[_sP( \overline{g})\sigma -\sigma (\overline{g})_sP]+\frac
12[Q(\overline{g})\omega -\omega (\overline{g})Q)]+ \\ 
+\frac 1{n-1}(\theta _1.\omega +\theta .Q)+\nabla _uQ+\frac 12[_sP( 
\overline{g})\omega +\omega (\overline{g})_sP]+ \\ 
+\frac 12[Q( \overline{g})\sigma +\sigma (\overline{g})Q]+ \\ 
+\frac 1{2e}[h_u(a)\otimes (g(u))(m+q)h_u-h_u((g(u))(m+q))\otimes h_u(a)]+
\\ 
+\frac 1e[Q(a)\otimes g(u)-g(u)\otimes Q(a)]+ \\ 
+\frac 12[h_u(\nabla _u\overline{g})_sP-_sP(\nabla _u\overline{g})h_u]+\frac
12[h_u(\nabla _u\overline{g})Q+Q(\nabla _u\overline{g})h_u]\text{ .}
\end{array}
\end{equation}

In index form 
\begin{equation}  \label{A.25}
(_TW_0)_{ij}=(_FW_0)_{ij}-(W_0)_{ij}\text{ .}
\end{equation}

n) Shear acceleration tensor, induced by the curvature, $_sM=M-\frac
1{n-1}.I.h_u$ 
\begin{equation}  \label{A.26}
\begin{array}{c}
M=\frac 1e.h_u(a)\otimes h_u(a)+\frac 12[_sE( \overline{g})\sigma +\sigma (%
\overline{g})_sE]+\frac 12[S(\overline{g})\omega +\omega (\overline{g})S]+
\\ 
+\frac 1{n-1}(\theta _o.\sigma +\theta ._sE)+\frac 1{n-1}(\theta
_o^{.}+\frac 1{n-1}.\theta _o.\theta )h_u+\nabla _u(_sE)+ \\ 
+\frac 12[_sE( \overline{g})\omega -\omega (\overline{g})_sE]+\frac 12[S(%
\overline{g})\sigma -\sigma (\overline{g})S]+ \\ 
+\frac 1{2e}[h_u(a)\otimes (g(u))(k_0+k-\nabla _u \overline{g}%
)h_u+h_u((g(u))(k_0+k-\nabla _u\overline{g}))\otimes h_u(a)]+ \\ 
+\frac 1e[_sE(a)\otimes g(u)+g(u)\otimes _sE(a)]+ \\ 
+\frac 12[h_u(\nabla _u \overline{g})_sE+_sE(\nabla _u\overline{g}%
)h_u]+\frac 12[h_u(\nabla _u \overline{g})S-S(\nabla _u\overline{g})h_u]- \\ 
-h_u(b_s)h_u\text{ .}
\end{array}
\end{equation}

In index form 
\begin{equation}  \label{A.27}
\begin{array}{c}
M_{ij}=M_{ji}=\frac 1e.h_{i \overline{k}}a^ka^lh_{\overline{l}%
j}+_sE_{k(i}\sigma _{j)l}g^{\overline{k} \overline{l}}+S_{k(i}\omega
_{j)l}g^{\overline{k}\overline{l}}+ \\ 
+\frac 1{n-1}(\theta _o.\sigma _{ij}+\theta ._sE_{ij})+\frac 1{n-1}(\theta
_o^{.}+\frac 1{n-1}.\theta _o.\theta )h_{ij}- \\ 
-_sE_{k(i}\omega _{j)l}g^{\overline{k}\overline{l}}-S_{k(i}\sigma _{j)l}g^{%
\overline{k}\overline{l}}+_sE_{ij;k}u^k+ \\ 
+\frac 1e.a^k[_sE_{i \overline{k}}u_j+_sE_{j\overline{k}}u_i+h_{\overline{k}%
(i}h_{j)\overline{l}}g^{ml}(e_{,m}-u_{\overline{n}}T_{mr}^nu^r- \\ 
-g_{rs;m}u^{\overline{r}}u^{\overline{s}}+g_{\overline{m}\overline{r}%
;s}u^su^r)]+ \\ 
+_sE_{\overline{k}(i}h_{j)\overline{l}}g^{kl}\text{ }_{;s}u^s+S_{\overline{k}%
(i}h_{j)\overline{l}}g^{kl}\text{ }_{;s}u^s- \\ 
-h_{\overline{k}(i}h_{j) \overline{l}}a^k\text{ }_{;m}g^{ml}\text{ .}
\end{array}
\end{equation}

o) Expansion acceleration, induced by the curvature, $I$%
\begin{equation}  \label{A.28}
\begin{array}{c}
I=-g[b]+ \overline{g}[_sE(\overline{g})\sigma ]+\overline{g}[S(\overline{g}%
)\omega ]+\theta _o^{.}+\frac 1{n-1}.\theta _o.\theta + \\ 
+\frac 1e[2g(u,\nabla _au)-g(u,T(a,u))+u(g(u,a))]- \\ 
-\frac 1{e^2}.g(u,a)[3g(u,a)+(\nabla _ug)(u,u)]\text{ .}
\end{array}
\end{equation}

In index form 
\begin{equation}  \label{A.29}
\begin{array}{c}
I=R_{ij}u^iu^j=-a^j \text{ }_{;j}+g^{\overline{i}\overline{j}}g^{\overline{k}%
\overline{l}}\text{ }_sE_{ik}\sigma _{lj}+g^{\overline{i}\overline{j}}g^{%
\overline{k}\overline{l}}S_{ik}\omega _{lj}+ \\ 
+\theta _o^{.}+\frac 1{n-1}.\theta _o.\theta +\frac 1e[a^k(e_{,k}-u_{%
\overline{n}}T_{km}^nu^m-g_{mn;k}u^{\overline{m}}u^{\overline{n}}- \\ 
-g_{\overline{k}\overline{m};s}u^su^m)+\frac 12(u^ke_{,k})_{,l}u^l-\frac
12.(g_{mn;r}u^r)_{;s}u^su^{\overline{m}}u^{\overline{n}}]- \\ 
-\frac 1{e^2}[\frac 34(e_{,k}u^k)^2-(e_{,k}u^k)g_{mn;r}u^ru^{\overline{m}}u^{%
\overline{n}}+\frac 14(g_{mn;r}u^ru^{\overline{m}}u^{\overline{n}})^2]\text{
.}
\end{array}
\end{equation}

p) Rotation expansion tensor, induced by the curvature, $N$%
\begin{equation}  \label{A.30}
\begin{array}{c}
N=\frac 12[_sE( \overline{g})\sigma -\sigma (\overline{g})_sE]+\frac 12[S(%
\overline{g})\omega -\omega (\overline{g})S]+ \\ 
+\frac 1{n-1}(\theta _o.\omega +\theta .S)+\nabla _uS+\frac 12[_sE( 
\overline{g})\omega +\omega (\overline{g})_sE]+ \\ 
+\frac 12[S( \overline{g})\sigma +\sigma (\overline{g})S]+ \\ 
+\frac 1{2e}[h_u(a)\otimes (g(u))(k_0+k-\nabla _u \overline{g}%
)h_u-h_u((g(u))(k_0+k-\nabla _u\overline{g}))\otimes h_u(a)]+ \\ 
+\frac 1e[S(a)\otimes g(u)-g(u)\otimes S(a)]+ \\ 
+\frac 12[h_u(\nabla _u \overline{g})_sE-_sE(\nabla _u\overline{g}%
)h_u]+\frac 12[h_u(\nabla _u \overline{g})S+S(\nabla _u\overline{g})h_u]- \\ 
-h_u(b_a)h_u\text{ .}
\end{array}
\end{equation}

In index form 
\begin{equation}  \label{A.31}
\begin{array}{c}
N_{ij}=-N_{ji}=_sE_{k[i}\sigma _{j]l}g^{\overline{k}\overline{l}%
}+S_{k[i}\omega _{j]l}g^{\overline{k}\overline{l}}+\frac 1{n-1}(\theta
_o.\omega _{ij}+\theta .S_{ij})- \\ 
-_sE_{k[i}\omega _{j]l}g^{\overline{k}\overline{l}}-S_{k[i}\sigma _{j]l}g^{%
\overline{k}\overline{l}}+S_{ij;k}u^k-h_{\overline{k}[i}h_{j]\overline{l}}a^k%
\text{ }_{;m}g^{ml}+ \\ 
+\frac 1e.a^k[S_{i \overline{k}}u_j-S_{j\overline{k}}u_i+h_{\overline{k}[%
i}h_{j]\overline{l}}g^{ml}(e_{,m}-u_{\overline{n}}T_{mr}^nu^r- \\ 
-g_{rs;m}u^{\overline{r}}u^{\overline{s}}+g_{\overline{m}\overline{r}%
;s}u^su^r)]-_sE_{\overline{k}[i}h_{j]\overline{l}}g^{kl}\text{ }_{;s}u^s-S_{%
\overline{k}[i}h_{j]\overline{l}}g^{kl}\text{ }_{;s}u^s\text{ .}
\end{array}
\end{equation}

\appendix

\section{Table 1. Kinematic characteristics connected with the notions
relative velocity and relative acceleration}

\subsection{Kinematic characteristics connected with the relative velocity:
........}

1. Relative position vector field

(relative position vector) .............................. $\xi _{\perp }= 
\overline{g}(h_u(\xi ))$

2. Relative velocity ...................................... $_{rel}v= 
\overline{g}(h_u(\nabla _u\xi ))$

3. Deformation velocity tensor

(deformation velocity, deformation) .. $d=d_0-d_1=\sigma +\omega +\frac
1{n-1}.\theta .h_u$

4. Torsion-free deformation velocity tensor

(torsion-free deformation velocity, torsion-free deformation)

.................................................................... $%
d_0=_sE+S+\frac 1{n-1}.\theta _o.h_u$

5. Deformation velocity tensor induced by the torsion

(torsion deformation velocity, torsion deformation)

.................................................................... $%
d_1=_sP+Q+\frac 1{n-1}.\theta _1.h_u$

6. Shear velocity tensor

(shear velocity, shear) .................................. $\sigma =_sE-_sP$

7. Torsion-free shear velocity tensor

(torsion shear velocity, torsion shear) ............ $_sE=E-\frac
1{n-1}.\theta _o.h_u$

8. Shear velocity tensor induced by the torsion

(torsion shear velocity tensor, torsion shear velocity, torsion shear)

..................................................................... $%
_sP=P-\frac 1{n-1}.\theta _1.h_u$

9. Rotation velocity tensor

(rotation velocity, rotation) ................................ $\omega =S-Q$

10. Torsion-free rotation velocity tensor

(torsion-free rotation velocity, torsion-free rotation)

......................................................................... $%
S=h_u(s)h_u$

11. Rotation velocity tensor induced by the torsion

(torsion rotation velocity, torsion rotation) .......... $Q=h_u(q)h_u$

12. Expansion velocity

(expansion) ....................................................... $\theta
=\theta _o-\theta _1$

13. Torsion-free expansion velocity

(torsion-free expansion) ..................................... $\theta _o= 
\overline{g}[E]$

14. Expansion velocity induced by the torsion

(torsion expansion velocity, torsion expansion) .... $\theta _1=\overline{g}[%
P]$ .

\subsection{Kinematic characteristics connected with the relative
acceleration:}

1. Acceleration ......................................................... $%
a=\nabla _uu$

2. Relative acceleration ................... ...................... $%
_{rel}a= $ $\overline{g}(h_u(\nabla _u\nabla _u\xi ))$

3. Deformation acceleration tensor

(deformation acceleration) ................................ $A=_sD+W+\frac
1{n-1}.U.h_u$

......................................................................... $%
A=A_0+G$

......................................................................... $%
A=_FA_0-_TA_0+G$

4. Torsion-free and curvature-free deformation acceleration tensor

(torsion-free and curvature-free deformation acceleration)

................................................................ $%
_FA_0=_{sF}D_0+_FW_0+\frac 1{n-1}._FU_0.h_u$

4.a. Curvature-free deformation acceleration tensor

(curvature-free deformation acceleration) ... $A_0=_sD_0+W_0+\frac
1{n-1}.U_0.h_u$

5. Deformation acceleration tensor induced by the torsion

(torsion deformation acceleration tensor, torsion deformation acceleration)

.................................................................$%
_TA_0=_{sT}D_0+_TW_0+\frac 1{n-1}._TU_0.h_u$

5.a. Deformation acceleration tensor induced by the curvature

(curvature deformation acceleration tensor, curvature deformation
acceleration)

.................................................................. $%
G=_sM+N+\frac 1{n-1}.I.h_u$

6. Shear acceleration tensor

(shear acceleration) ................................... $_sD=D-\frac
1{n-1}.U.h_u$

.................................................................. $%
_sD=_sD_0+_sM$

.................................................................. $%
_sD=_{sF}D_0-_{sT}D_0+_sM$

7. Torsion-free and curvature-free shear acceleration tensor

(torsion-free and curvature-free shear acceleration)

................................................................. $%
_{sF}D_0=_FD_0-\frac 1{n-1}._FU_0.h_u$

7.a. Curvature-free shear acceleration tensor

(curvature-free shear acceleration) .............. $_sD_0=D_0-\frac
1{n-1}.U_0.h_u$

................................................................... $%
_sD_0=_{sF}D_0-_{sT}D_0$

8. Shear acceleration tensor induced by the torsion

(torsion shear acceleration tensor, torsion shear acceleration)

.................................................................. $%
_{sT}D_0=_TD_0-\frac 1{n-1}._TU_0.h_u$

8.a. Shear acceleration tensor induced by the curvature

(curvature shear acceleration tensor, curvature shear acceleration)

.................................................................... $%
_sM=M-\frac 1{n-1}.I.h_u$

9. Rotation acceleration tensor

(rotation acceleration) .................................... $W=W_0+N$

...................................................................... $%
W=_FW_0-_TW_0+N$

10. Torsion-free and curvature-free rotation acceleration tensor

(torsion-free and curvature-free rotation acceleration)

.................................................................... $%
_FW_0=h_u(b_a)h_u$

10.a. Curvature-free rotation acceleration tensor

(curvature-free rotation acceleration) .............. $W_0=W-N$

...................................................................... $%
W_0=_FW_0-_TW_0$

11. Rotation acceleration tensor induced by the torsion

(torsion rotation acceleration tensor, torsion rotation acceleration)

.................................................................... $%
_TW_0=_FW_0-W_0$

11.a. Rotation acceleration tensor induced by the curvature

(curvature rotation acceleration tensor, curvature rotation acceleration)

...................................................................... $%
N=h_u(K_a)h_u$

12. Expansion acceleration ............................. $U=U_0+I$

...................................................................... $%
U=_FU_0-_TU_0+I$

13. Torsion-free and curvature-free expansion acceleration

.................................................................... $_FU_0= 
\overline{g}[_FD_0]$

13.a. Curvature-free expansion acceleration ... $U_0=\overline{g}[D_0]$

.......................................................................$%
U_0=_FU_0-_TU_0$

14. Expansion acceleration induced by the torsion

(torsion expansion acceleration) .................. ... $U_0=\overline{g}[%
_TD_0]$

14.a. Expansion acceleration induced by the curvature

(curvature expansion acceleration) ................... $I=\overline{g}[M]= 
\overline{g}[G]$

\section{Table 2. Classification of non-isotropic auto-parallel vector
fields on the basis of the kinematic characteristics connected with the
relative velocity and relative acceleration}

\subsection{Classification on the basis of kinematic characteristics
connected with the relative velocity}

The following conditions, connected with the relative velocity, can
characterize the vector fields over manifolds with affine connection and
metric:

1. $\sigma =0$.

2. $\omega =0$.

3. $\theta =0$.

4. $\sigma =0$, $\omega =0$.

5. $\sigma =0$, $\theta =0$.

6. $\omega =0$, $\theta =0$.

7. $\sigma =0$, $\omega =0$, $\theta =0$.

8. $_sE=0$.

9. $S=0$.

10. $\theta _o=0$.

11. $_sE=0$, $S=0$.

12. $_sE=0$, $\theta _o=0$.

13. $S=0$, $\theta _o=0$.

14. $_sE=0$, $S=0$, $\theta _o=0$.

15. $_sP=0.$

16. $Q=0$.

17. $\theta _1=0$.

18. $_sP=0$, $Q=0$.

19. $_sP=0$, $\theta _1=0$.

20. $Q=0$, $\theta _1=0$.

21. $_sP=0$, $Q=0$, $\theta _1=0$.

\subsection{Classification on the basis of the kinematic characteristics
connected with the relative acceleration}

The following conditions, connected with the relative acceleration, can
characterize the vector fields over manifolds with affine connection and
metric:

1. $_sD=0$.

2. $W=0$.

3. $U=0$.

4. $_sD=0$, $W=0$.

5. $_sD=0$, $U=0$.

6. $W=0$, $U=0$.

7. $_sD=0$, $W=0$, $U=0$.

8. $_sM=0$.

9. $N=0$.

10. $I=0$.

11. $_sM=0$, $N=0$.

12. $_sM=0$, $I=0$.

13. $N=0$, $I=0$.

14. $_sM=0$, $N=0$, $I=0$.

15. $_{sT}D_0=0$.

16. $_TW_0=0$.

17. $_TU_0=0$.

18. $_{sT}D_0=0$, $_TW_0=0$.

19. $_{sT}D_0=0$, $_TU_0=0$.

20. $_TW_0=0$, $_TU_0=0$.

21. $_{sT}D_0=0$, $_TW_0=0$, $_TU_0=0$.

\end{document}